\documentclass[12pt]{article}
\usepackage{amscd,amsmath,amssymb,amstext,amsthm,exscale,latexsym}
\usepackage{graphicx}
\newcommand {\al}   {\alpha}       \newcommand {\bt}  {\beta}

\newcommand {\lm}   {\lambda}      
          
\newcommand {\s }   {\sigma}       
         
\newcommand {\vf }  {\varphi}      
         
\newcommand {\Lm}   {\Lambda}      \newcommand {\Om}  {\Omega}
       
\newcommand {\pl}   {\partial}     \newcommand {\nb}  {\nabla}
       \renewcommand {\cos}{{\sf\,cos\,}}

\renewcommand {\det}{{\sf\,det\,}}

       \renewcommand {\lim}{{\sf\,lim\,}}

\newcommand   {\sign}{{\sf\,sign\,}}

\newcommand   {\const}{{\sf\,const}}

\newcommand {\MI}  {{\mathbb I}}   
   \newcommand {\ML}  {{\mathbb L}}
\newcommand {\MM}  {{\mathbb M}}   
\newcommand {\MO}  {{\mathbb O}}   
   \newcommand {\MR}  {{\mathbb R}}
\newcommand {\MS}  {{\mathbb S}}   
\newcommand {\MU}  {{\mathbb U}}   \newcommand {\MV}  {{\mathbb V}}

\newcommand {\Si}  {{\textsc{i}}}   
   \newcommand {\Sl}  {{\textsc{l}}}

\newtheorem{prop}{Proposition}[section]
\newtheorem{theorem}{Theoremа}[section]

\theoremstyle{definition}
\begin{document}
\title     {Global properties of warped solutions in General Relativity with an
            electromagnetic field and a cosmological constant. II.}
\author    {D. E. Afanasev
            \thanks{E-mail: daniel\_afanasev@yahoo.com}\\
            \sl High school N1561, ul.\ Paustovskogo, 6, kor.\ 2, 117464,
            \sl Moscow\\
            M. O. Katanaev
            \thanks{E-mail: katanaev@mi-ras.ru}\\
            \sl Steklov mathematical institute,
            \sl ul.~Gubkina, 8, Moscow, 119991, Russia}
\date      {20 March 2019}
\maketitle
\begin{abstract}
We consider general relativity with cosmological constant minimally coupled to
the electromagnetic field and assume that the four-dimensional space-time
manifold is a warped product of two surfaces with Lorentzian and Euclidean
signature metrics. Field equations imply that at least one of the surfaces
must be of constant curvature leading to the symmetry of the metric
(``spontaneous symmetry emergence''). We classify all global solutions in the
case when the Lorentzian surface is of constant curvature (case {\sf C}). These
solutions are invariant with respect to the Lorentz
${\mathbb S}{\mathbb O}(1,2)$ or Poincare $\MI\MO(1,1)$ groups acting on the
Lorentzian surface.
\end{abstract}
\section{Introduction}
In this paper, we continue our work ref.\ \cite{AfaKat19}.

Physical interpretation of solutions to Einstein's equations relies on the
knowledge of the global space-time structure, that is we must know not only a
solution to Einstein's equations in some coordinate chart but the space-time
itself must be maximally extended along geodesic lines. It means that any
geodesic can be either continued to infinite value of the canonical parameter in
both directions or it ends up at a singular point where one of the geometric
invariants, for example, scalar curvature, becomes singular.
There are many well known exact solutions in general relativity (see, e.g.\
\cite{KrStMaHe80,GriPod09}) but only part of them are analyzed globally.
The famous example is the Kruskal--Szekeres extension \cite{Kruska60,Szeker60}
of the Schwarzschild solution. In this case, the space-time is globally the
topological product of a sphere (spherical symmetry) with the two-dimensional
Lorentzian surface depicted by the well known Carter--Penrose diagram. Precisely
this knowledge of the global structure allows one to introduce the notion of
black and white holes.

The Reissner--Nordstr\"om solution \cite{Reissn16,Nordst18} is the spherically
symmetric solution of Eins\-tein's equations with the electromagnetic field and
depends on two parameters: mass and charge. Depending on the relation between
mass and charge, there are three global solutions: the Reissner-Nordstr\"om
black hole, extremal black hole, and naked singularity. The spherically
symmetric exact solution of Einstein's equations with electromagnetic field and
cosmological constant is known locally and depends on three parameters:
mass, charge, and cosmological constant. For some values of these parameters the
space-time was known globally. The detailed analysis of all global solutions in
this case was given in \cite{AfaKat19}. They enter the cases {\sf A} and
{\sf B}.

The case {\sf A} consists of global solutions, which are the product of two
constant curvature surfaces. Solutions of the form of the warped product of
constant curvature Riemannian surface with some Lorentzian surface depicted by
the Carter--Penrose diagram constitute the case {\sf B}. It includes three
subcases corresponding to three possible Riemannian constant curvature surfaces:
two-dimensional sphere (spherically symmetric solutions), Euclidean plane
(planar solutions), and two-sheeted hyperboloid ($\MS\MO(1,2)$-symmetric
solutions). In this paper, we construct all global solutions when the space-time
is the warped product of the Lorentzian surface of constant curvature
(one-sheeted hyperboloid or Minkowskian plane) and a Riemannian surface
(Case {\sf C}).  The solutions are classified for all values of three
parameters: mass, charge, and cosmological constant. Totally, we have found 19
global solutions in case {\sf C}.
The cases {\sf A, B,} and {\sf C} exhaust all possible solutions having the form
of the warped product of two surfaces.

In general, some four-dimensional spacetimes may have more then one
representation in the form of the product of two surfaces. It means that some
of the solutions found in the paper may topologically coincide. Analysis of
these possibilities requires a different techniques and is out of the scope of
the present investigation.

We do not assume that solutions have any symmetry from the very beginning.
Instead, we require the space-time to be the warped product of two surfaces,
${\mathbb M}={\mathbb U}\times{\mathbb V}$, where ${\mathbb U}$ and
${\mathbb V}$ are two two-dimensional surfaces with Lorentzian and Euclidean
signature metrics, respectively. As the consequence of the field equations,
at least one of the surfaces must be of constant curvature. In this paper, we
consider the case when the surface ${\mathbb U}$ is of constant curvature (case
{\sf C}). There are two possibilities: ${\mathbb U}$ is the one-sheeted
hyperboloid ${\mathbb H}^2$ (the Lorentzian ${\mathbb S}{\mathbb O}(1,2)$
symmetry), or the Minkowskian plane $\MR^{1,1}$ (the Poincare
${\mathbb I}{\mathbb S}{\mathbb O}(1,1)$ symmetry). We see that the symmetry of
solutions is not assumed from the beginning but arises as the consequence of the
field equations. This effect is called ``spontaneous symmetry emergence''.
We classify all global solutions by constructing explicitly all Riemannian
maximally extended surfaces ${\mathbb V}$ depending on relations between mass,
charge, and cosmological constant. Moreover, we prove that there is the
additional fourth Killing vector field in each case. This is a generalization of
Birkhoff's theorem.

The emergence of extra symmetry due to the field equations is known for a long
time: the famous Birkhoff theorem states that there is extra Killing vector
field in the spherically symmetric spacetime. This effect is called
``Birkhoff-like theorem'' in ref. \cite{Schmid13}. The term ``spontaneous
symmetry emergence'' seems to be more general and it is the counterpart to
``spontaneous symmetry breaking'' in gauge models.

Global structure of space-times in general relativity and 2d-gravity was
analyzed, e.g.\ in [9--13].
\nocite{Walker70,Carter73,Katana93A,KloStr96C,KloStr97}
For review, see \cite{GriPod09}. In particular, global planar and Lobachevsky
plane solutions in general relativity were described in
\cite{HuaLia95,Lemos95,BrLoPe97}. These studies were restricted to cases
{\sf A} and {\sf B} when the second multiplier $\MV$ in the product
$\MM=\MU\times\MV$ is of constant curvature. Classification of all global
solutions in cases {\sf A} and {\sf B} is given in \cite{AfaKat19}.
The analysis of case {\sf C} given in the present paper seems to be new.

The (2+2)-decomposition of spacetime is applicable not only to general
relativity, but to other gravity models as well. For example, some new vacuum
solutions were obtained in conformal Weyl gravity \cite{DzhSch00}.

This paper follows the classification of global warped product solutions of
general relativity with cosmological constant (without electromagnetic field)
given in \cite{KaKlKu99}. The maximally extended Riemannian surfaces with
one Killing vector field are constructed using the method described in
\cite{Katana09}.

As in \cite{KaKlKu99}, we assume that the space-time ${\mathbb M}$ is the warped
product of two surfaces, ${\mathbb M}={\mathbb U}\times{\mathbb V}$, where
${\mathbb U}$ and ${\mathbb V}$ are surfaces with Lorentzian and Euclidean
signature metrics, respectively. Local coordinates on ${\mathbb M}$ are denoted
by $\hat x^i$, $i=0,1,2,3$, and coordinates on the surfaces by Greek letters
from the beginning and the middle of the alphabet:
\begin{equation*}
  (x^\alpha)\in{\mathbb U},\quad\alpha=0,1,\qquad
  (y^\mu)\in{\mathbb V},\quad\mu=2,3.
\end{equation*}
That is $(\hat x^i):=(x^\alpha,y^\mu)$. Geometrical notions on four-dimensional
space-time are marked by the hat to distinguish them from the notions on
surfaces ${\mathbb U}$ and ${\mathbb V}$, which appear more often.

We do not assume any symmetry of solutions from the very beginning.

The four-dimensional metric of the warped product of two surfaces has block
diagonal form by definition:
\begin{equation}                                                  \label{egdtrf}
  \widehat g_{ij}=\begin{pmatrix} k(y)g_{\alpha\beta}(x) & 0 \\
  0 & m(x)h_{\mu\nu}(y)
  \end{pmatrix},
\end{equation}
where $g_{\alpha\beta}(x)$ and $h_{\mu\nu}(y)$ are some metrics on surfaces
${\mathbb U}$ and ${\mathbb V}$, respectively, $k(y)\ne0$ and $m(x)\ne0$ are
scalar (dilaton) fields on ${\mathbb V}$ and ${\mathbb U}$. Without loss of
generality, signatures of two-dimensional metrics $g_{\alpha\beta}$ and
$h_{\mu\nu}$ are assumed to be $(+-)$ and $(++)$ or $(--)$, respectively. In the
rigorous sense, the metric (\ref{egdtrf}) is the doubly warped product. It
reduces to the usual warped product for $k=\text{const}$ or $m=\text{const}$.

The Ricci tensor and the scalar curvature for metric (\ref{egdtrf}) are computed
in \cite{AfaKat19,KaKlKu99}.
\section{Solution for the electromagnetic field}
In this and the next section, we shortly review local solution of the equations
of motion \cite{AfaKat19} for convenience of a reader.

We consider the following action
\begin{equation}                                                  \label{ubcftg}
  S=\int\!d^4x\sqrt{|\widehat g|}
  \left(\widehat R-2\Lambda-\frac14\widehat F^2\right),
\end{equation}
where $\widehat R$ is the four-dimensional scalar curvature for metric
$\widehat g_{ij}$, $\widehat g:=\det\widehat g_{ij}$, $\Lambda$ is a
cosmological constant, and $\widehat F^2$ is the square of the electromagnetic
field strength:
\begin{equation*}
  \widehat F^2:=\widehat F_{ij}\widehat F^{ij},\qquad
  \widehat F_{ij}:=\partial_i \widehat A_j-\partial_j\widehat A_i,
\end{equation*}
for the electromagnetic field potential $\widehat A_i$.

To simplify the problem, we assume that the four-dimensional electromagnetic
potential consists of two parts:
\begin{equation*}
  \widehat A_i:=\big(A_\alpha(x),A_\mu(y)\big),
\end{equation*}
where $A_\alpha(x)$ and $A_\mu(y)$ are two-dimensional electromagnetic potentials
on surfaces ${\mathbb U}$ and ${\mathbb V}$, respectively. Then the
electromagnetic field strength becomes block diagonal
\begin{equation}                                                  \label{unfhgt}
  \widehat F_{ij}=\begin{pmatrix}F_{\alpha\beta} & 0 \\ 0 & F_{\mu\nu}
  \end{pmatrix},
\end{equation}
where
\begin{equation*}
  F_{\alpha\beta}(x):=\partial_\alpha A_\beta-\partial_\beta A_\alpha,\qquad
  F_{\mu\nu}(y):=\partial_\mu A_\nu-\partial_\nu A_\mu
\end{equation*}
are strength components for the two-dimensional electromagnetic potentials.

Variation of the action (\ref{ubcftg}) with respect to the metric yields
four-dimensional Einstein's equations:
\begin{equation}                                                  \label{ubsvgr}
  \widehat R_{ij}-\frac12\widehat g_{ij}\widehat R+\widehat g_{ij}\Lambda
  =-\frac12 \widehat T_{{\textsc{e}}{\textsc{m}} ij},
\end{equation}
where
\begin{equation}                                                  \label{ubnhgy}
  \widehat T_{{\textsc{e}}{\textsc{m}} ij}:=-\widehat F_{ik}\widehat F_j{}^k
  +\frac14\widehat g_{ij}\widehat F^2
\end{equation}
is the electromagnetic field energy-momentum tensor. Variation of
the action with respect to electromagnetic field yields Maxwell's equations:
\begin{equation}                                                  \label{uncmhy}
  \partial_j\big(\sqrt{|\widehat g|}\widehat F^{ji}\big)=0,
\end{equation}
where
\begin{equation*}
  \widehat g=k^2m^2gh,\qquad g:=\det g_{\alpha\beta},\qquad h:=\det h_{\mu\nu}.
\end{equation*}

In what follows, the raising of Greek indices from the beginning and the middle
of the alphabet is performed by using the inverse metrics $g^{\alpha\beta}$ and
$h^{\mu\nu}$. Therefore
\begin{equation*}
  \widehat F^2=\frac1{k^2}F_{\alpha\beta}F^{\alpha\beta}
  +\frac1{m^2}F_{\mu\nu}F^{\mu\nu}.
\end{equation*}

In the case under consideration, Maxwell's Eqs.~(\ref{uncmhy}) for $i=\alpha$
lead to the equality
\begin{equation*}
 \frac1{|k|}\sqrt{|h|}\partial_\beta\left(|m|\sqrt{|g|}F^{\beta\alpha}\right)=0.
\end{equation*}
A general solution to these equations has the form
\begin{equation}                                                  \label{ubvcfr}
  F^{\alpha\beta}=\frac{2Q}{|m|}\varepsilon^{\alpha\beta},\qquad Q=\const,
\end{equation}
where $Q$ -- is a constant of integration (electromagnetic charge) and
$\varepsilon^{\alpha\beta}$ is the totally antisymmetric second rank
tensor. The factor 2 in the right hand side is introduced for simplification of
subsequent formulae.

If $i=\mu$, then Maxwell's Eqs.~(\ref{uncmhy}) yield the equality
\begin{equation*}
  \frac1{|m|}\sqrt{|g|}\partial_\mu\left(|k|\sqrt h F^{\mu\nu}\right)=0.
\end{equation*}
Its general solution is
\begin{equation}                                                  \label{unbgtm}
  F^{\mu\nu}=\frac{2P}{|k|}\varepsilon^{\mu\nu},\qquad P={\sf\,const}.
\end{equation}

Now the four-dimensional electromagnetic energy-momentum tensor (\ref{ubnhgy})
becomes:
\begin{equation}                                                  \label{uvbxfp}
  \widehat T_{ij}=\begin{pmatrix}
  \widehat T_{\alpha\beta} & 0 \\ 0 & \widehat T_{\mu\nu} \end{pmatrix},
\end{equation}
where
\begin{equation*}
  \widehat T_{\alpha\beta}=\frac{2g_{\alpha\beta}}{km^2}(Q^2+P^2),\qquad
  \widehat T_{\mu\nu}=-\frac{2h_{\mu\nu}}{k^2m}(Q^2+P^2).
\end{equation*}

Now we have to solve Einstein's Eqs.~(\ref{ubsvgr}) with the right hand side
(\ref{uvbxfp}). Since energy-momentum tensor depends only on the sum $Q^2+P^2$,
we set $P=0$ to simplify formulae. In the final answer, this constant is easily
reconstructed by substitution $Q^2\mapsto Q^2+P^2$.

In what follows, we consider only the case $Q\ne0$, because the case $Q=0$ was
considered in \cite{KaKlKu99} in full detail.
\section{Einstein's equations}
The right hand side of Einstein's Eqs.~(\ref{ubsvgr}) is defined by the general
solution of Maxwell's equations, which leads to the electromagnetic
energy-momentum tensor (\ref{uvbxfp}). The trace of Einstein's equations is
easily solved with respect to the scalar curvature: $\widehat R=4\Lambda$. After
elimination of the scalar curvature, Einstein's equations simplify
\begin{equation}                                                  \label{ubbdgt}
  \widehat R_{ij}-\widehat g_{ij}\Lambda=-\frac12\widehat
  T_{{\textsc{e}}{\textsc{m}} ij}.
\end{equation}
For indices $(ij)=(\alpha,\beta)$, $(\mu\nu)$, and $(\alpha,\mu)$, these
equations yield the following system:
\begin{align}                                                     \label{ubvcfl}
  R_{\alpha\beta}+\frac{\nabla_\alpha\nabla_\beta m}m
  -\frac{\nabla_\alpha m\nabla_\beta m}{2m^2}
  +g_{\alpha\beta}\left(\frac{\nabla^2k}{2m}-k\Lambda
  +\frac{Q^2}{m^2k}\right)=&0,
\\                                                                \label{ubncjh}
  R_{\mu\nu}+\frac{\nabla_\mu\nabla_\nu k}k
  -\frac{\nabla_\mu k\nabla_\nu k}{2k^2}
  +h_{\mu\nu}\left(\frac{\nabla^2 m}{2k}-m\Lambda-\frac{Q^2}{k^2m}\right)=&0,
\\
  -\frac{\nabla_\alpha m\nabla_\mu k}{2mk}=&0,
\end{align}
where $R_{\alpha\beta}$ and $R_{\mu\nu}$ are the Ricci tensors for
two-dimensional metrics $g_{\alpha\beta}$ and $h_{\mu\nu}$, respectively,
$\nabla_\alpha$ and $\nabla_\mu$ are two-dimensional covariant derivatives with
Christoffel's sym\-bols on surfaces ${\mathbb U}$ and ${\mathbb V}$,
$\nabla^2:=g^{\alpha\beta}\nabla_\alpha\nabla_\beta$ or
$\nabla^2:=h^{\mu\nu}\nabla_\mu\nabla_\nu$, which is clear from the context.

The full system of Einstein's equations after extracting the traces from
Eqs.\ (\ref{ubvcfl}) and (\ref{ubncjh}) takes the form
\begin{align}                                                     \label{unncbg}
  \nabla_\alpha\nabla_\beta m-\frac{\nabla_\alpha m\nabla_\beta m}{2m}
  -\frac12\left(\nabla^2 m
  -\frac{(\nabla m)^2}{2m}\right)=&0,
\\                                                                \label{ubmsdi}
  \nabla_\mu\nabla_\nu k-\frac{\nabla_\mu k\nabla_\nu k}{2k}
  -\frac12\left(\nabla^2 k
  -\frac{(\nabla k)^2}{2k}\right)=&0,
\\                                                                \label{undbyt}
  R^{(g)}+\frac{\nabla^2 m}m-\frac{(\nabla m)^2}{2m^2}
  +\frac{\nabla^2 k}m-2k\Lambda
  +\frac{2Q^2}{m^2k}=&0,
\\                                                                \label{undhtt}
  R^{(h)}+\frac{\nabla^2 k}k-\frac{(\nabla k)^2}{2k^2}
  +\frac{\nabla^2 m}k-2m\Lambda
  -\frac{2Q^2}{k^2m}=&0,
\\                                                                \label{ubvfds}
  \nabla_\alpha m\nabla_\mu k=&0,
\end{align}
where $(\nabla m)^2:=g^{\alpha\beta}\nabla_\alpha m\nabla_\beta m$,
$(\nabla k)^2:=g^{\mu\nu}\nabla_\mu k\nabla_\nu k$, $R^{(g)}$ and $R^{(h)}$ are
scalar curvatures of two-dimensional surfaces ${\mathbb U}$ and ${\mathbb V}$
for metrics $g$ and $h$, respectively.

The last Eq.~(\ref{ubvfds}), which corresponds to mixed values of indices
$(ij)=(\alpha\mu)$ in Einstein's equations, results in strong restrictions on
solutions. Namely, there are only three cases:
\begin{equation}                                                  \label{ecasek}
\begin{array}{lrr}
  {\sf A}:  & \qquad k={\sf\,const}\ne0,      & \qquad m={\sf\,const}\ne0, \\
  {\sf B}:  & k={\sf\,const}\ne0,      & \nabla_\alpha m\ne0, \\
  {\sf C}:  & \nabla_\mu k\ne0,  & m={\sf\,const}\ne0.
\end{array}
\end{equation}
The cases {\sf A} and {\sf B} were considered in \cite{AfaKat19}. Now we
consider the third case {\sf C} in detail.
\section{Lorentz-invariant and planar solutions                 \label{smlcos}}
In this section, we consider the case {\sf C} (\ref{ecasek}) when the second
dilaton field included in the warped product (\ref{egdtrf}) is constant,
$m=\const$. As will be shown below, the Lorentzian surface $\MU$ must be of
constant curvature in this case. Therefore, it can be the one-sheeted
hyperboloid, $\MU=\ML^2$, for $R^{(g)}=\const\ne0$, or its universal covering.
Then global solutions of Einstein's equations have the form of the topological
product $\MM=\ML^2\times\MV$. The surface $\MU$ can also be the Minkowskian
plane $\MR^{1,1}$ or its factor-spaces for $R^{(g)}=0$. This global solution has
the form $\MR^{1,1}\times\MV$. We shall see that field equations imply that
the second factor $\MV$ has one more Killing vector.

The surface $\MV$ depends on relations between constants of integration and a
cosmological constant and can have conical singularities and/or singularities of
curvature along the edge of the surface, as in the absence of the
electromagnetic field \cite{KaKlKu99}.
From physical point of view, these singularities correspond to cosmic strings or
singular domain walls that evolve in time.

Without loss of generality, we fix $m=1$ and suppose that the metric
$h_{\mu\nu}$ can be either positive or negative definite. In both cases, the
signature of the four-dimensional metric will be Lorentzian: $(+---)$ or
$(+-++)$.

The solution of equations (\ref{unncbg})--(\ref{ubvfds}) is carried out similar
to the case {\sf B}: we have only to put $m=1$. Therefore we briefly describe
the main steps of the calculations, emphasizing the moments that are specific to
the Euclidean signature.

For $m=1$ the complete system of Einstein's equations
(\ref{unncbg})--(\ref{ubvfds}) takes the form
\begin{align}                                                     \label{qhngtr}
  \nb_\mu\nb_\nu k-\frac{\nb_\mu k\nb_\nu k}{2k}-\frac12h_{\mu\nu}
  \left[\nb^2k-\frac{(\nb k)^2}{2k}\right]&=0,
\\                                                                \label{qfrjds}
  R^{(g)}+\nb^2 k-2k\Lm+\frac{2Q^2}k&=0,
\\                                                                \label{qgedas}
  R^{(h)}+\frac{\nb^2 k}k-\frac{(\nb k)^2}{2k^2}-2\Lm-\frac{2Q^2}{k^2}&=0.
\end{align}
As in the case of {\sf B}, equation (\ref{qfrjds}) includes the sum of
functions of different arguments: $R^{(g)}=R^{(g)}(x)$ and $k=k(y)$. Therefore
the scalar curvature of surface $\MU$ must be constant,
$R^{(g)}=-2K^{(g)}=\const$. It implies that the surface $\MU$ is the
one-sheeted hyperboloid $\ML^2$ or its universal covering for $K^{(g)}\ne0$.
If $K^{(g)}=0$, then the surface $\MU$ is the Minkowskian plane $\MR^{1,1}$ or
its factor spaces.

This is a very important consequence of Einstein's equations, because all
solutions for $K^{(g)}\ne0$ must be $\MO(1,2)$-invariant, the Lorentz
transformation group acting on the one-sheeted hyperboloid with coordinates
$x^0,x^1$.
Therefore global solutions of class ${\sf C}$ are called Lorentz-invariant. If
$K^{(g)}=0$, then the symmetry group is the Poincare group $\MI\MO(1,1)$. We see
that the symmetry of the metric arises from Einstein's equations. This is the
spontaneous symmetry emergence.

Then equation (\ref{qfrjds}) is written as
\begin{equation}                                                  \label{qnmasw}
  \nb^2 k-2\left(k\Lm+K^{(g)}-\frac{Q^2}k\right)=0,
\end{equation}
where $K^{(g)}=\const$.

The proof of the following statement is similar to the case {\sf B}.
\begin{prop}
Equation (\ref{qnmasw}) is the first integral of equations (\ref{qhngtr}) and
(\ref{qgedas}) for $\nb_\mu k\ne0$ on the whole $\MV$.
\end{prop}
Hence it is sufficient to solve only Eqs.\ (\ref{qhngtr}) and
(\ref{qnmasw}), Eq.\ (\ref{qgedas}) being their consequence.

The next step is to fix the coordinates on surface $\MV$. The conformally flat
Euclidean metric on surface $\MV$ is
\begin{equation}                                                  \label{ebogae}
  h_{\mu\nu}dy^\mu dy^\nu=\Phi dzd\bar z=\Phi(d\s^2+d\rho^2),\qquad\Phi\ne0.
\end{equation}
where the conformal factor $\Phi(z,\bar z)$ is the function of complex
coordinates:
\begin{equation}                                                  \label{ecbcov}
  z:=\s+i\rho,\qquad \bar z=\s-i\rho,
\end{equation}
where $\s=y^2$, $\rho=y^3$. The metric of the whole four-dimensional spacetime
is
\begin{equation}                                                  \label{einnoe}
  ds^2=kd\Om_\Sl+\Phi dzd\bar z,
\end{equation}
where $d\Om_\Sl$ is the metric of constant curvature on the one-sheeted
hyperboloid $\ML^2$ for $K^{(g)}\ne0$. It can be written, for example, in
stereographic coordinates
\begin{equation}                                                  \label{ecoclm}
  d\Om_\Sl=g_{\al\bt}dx^\al dx^\bt=\frac{dt^2-dx^2}
  {\left[1+\frac{K^{(g)}}4(t^2-x^2)\right]^2},
\end{equation}
where $x^0:=t$, $x^1:=x$.

Without loss of generality, we consider positive $k>0$. Otherwise we can
rearrange the first two coordinates $x^0$ and $x^1$ on $\MU$. Now we
introduce the parameterization
\begin{equation}                                                  \label{eshdgf}
  k=q^2,\qquad q>0.
\end{equation}
Then we have the following system of equations for two unknown functions $q$ and
$\Phi$ instead of equations (\ref{qhngtr}) and (\ref{qnmasw})
\begin{align}                                                     \label{eiuoha}
  \pl^2_{zz}q-\frac{\pl_z\Phi\pl_zq}\Phi&=0,
\\                                                                \label{eggohb}
  \pl^2_{\bar z\bar z}q-\frac{\pl_{\bar z}\Phi\pl_{\bar z}k}\Phi&=0,
\\                                                                \label{egcmhc}
  2\frac{\pl_z\pl_{\bar z}q^2}\Phi-\left(K^{(g)}+\Lm q^2
  -\frac{Q^2}{q^2}\right)&=0.
\end{align}
Similarly to the case {\sf B}, any solution of Eqs.\ (\ref{eiuoha}) and
(\ref{eggohb}) depends on the single variable: $q=q(z\pm\bar z)$ and
$\Phi=\Phi(z\pm\bar z)$, and the function $\Phi$ is determined by equation
\begin{equation}                                                  \label{ezintk}
  |\Phi|=|q'|,
\end{equation}
where the prime denotes differentiation with respect to the corresponding
argument. In the arguments $z\pm\bar z$, the lower and upper signs correspond to
positively and negatively definite Riemannian metric on $\MV$. Thus, the
functions $q$ and $\Phi$ depend either on coordinate $\s$, or on $i\rho$.
Both choices are equivalent because of the rotational $\MO(2)$-symmetry of the
Euclidean metric in Eq.~(\ref{ebogae}). Therefore, for definiteness, we
assume that functions $q(\s)$ and $\Phi(\s)$ depend on $\s$. The factor $i$
in the argument $i\rho$ does not matter, because equality (\ref{ezintk})
contains modules.

Then equation (\ref{egcmhc}) is
\begin{equation}                                                  \label{esecik}
  \frac12(q^2)''=\left(K^{(g)}+\Lm q^2-\frac{Q^2}{q^2}\right)\Phi,
\end{equation}
where the prime denotes differentiation with respect to $\s$. To integrate it,
we have to remove moduli signs in Eq.\ (\ref{ezintk}).

Let us consider the case of $\Phi q'>0$. Taking into account Eq.~(\ref{ezintk}),
Eq.\ (\ref{esecik}), takes the form
\begin{equation*}
  \frac12(q^2)''=\left(K^{(g)}+\Lm q^2-\frac{Q^2}{q^2}\right)q',
\end{equation*}
and is easily integrated
\begin{equation}                                                  \label{enfokw}
  q'=K^{(g)}-\frac{2M}q+\frac{Q^2}{q^2}+\frac{\Lm q^2}3,
\end{equation}
where $M\in\MR$ is an arbitrary integration constant. It is denoted as the mass
in the Schwarzschild solution but now it cannot have this interpretation. Using
equality (\ref{ezintk}), we obtain the conformal factor
\begin{equation}                                                  \label{efihkp}
  \Phi(q)=K^{(g)}-\frac{2M}q+\frac{Q^2}{q^2}+\frac{\Lm q^2}3
\end{equation}
This expression differs from the conformal factor in the spatially symmetric
case {\sf B} only by the sign of the cosmological constant.

The case $\Phi q'<0$ is integrated in the same way. Finally, the general
solution of Einstein's equations in case {\sf C} takes the form
\begin{equation}                                                  \label{emsdty}
  ds^2=q^2d\Om_\Sl+\Phi (q)(d\s^2+d\rho^2),
\end{equation}
where the conformal factor is given by Eq.~(\ref{efihkp}) and the function
$q=q(\s)$ is defined by Eq.~(\ref{ezintk}). Thus, the metric on surface $\MV$
has one Killing vector $\pl_\rho$. Therefore, we can use the algorithm for
construction of global Riemannian surfaces with one Killing vector field given
in \cite{Katana09} and briefly summarized in the next section.

Choosing the function $q(\s)$ as one of the coordinates, metric (\ref{emsdty})
can be written locally in the Schwarzschild-like form
\begin{equation}                                                  \label{egbsty}
  ds^2=q^2d\Om_\Sl+\frac{dq^2}{\Phi(q)}+\Phi (q)d\s^2.
\end{equation}

The resulting metric (\ref{egbsty}) has three Killing vectors, corresponding to
the symmetry group $\MO(1,2)$ of the one-sheeted hyperboloid of constant
curvature $\ML^2$, and one additional Killing vector $\pl_\rho$ on surface $\MV$
(the analogue of Birkhoff's theorem).

Equations (\ref{qgedas}), (\ref{eshdgf}), (\ref{ezintk}), and (\ref{efihkp})
imply the expression for the scalar curvature of surface $\MV$
$$
  R^{(h)}=\frac23\Lm-\frac{4M}{q^3}+\frac{6Q^2}{q^4}.
$$
It shows that the point $q=0$ is a true singularity for $M\ne0$ and/or $Q\ne0$,
and therefore the extension of surface $\MV$ is impossible through $q=0$.
\section{Riemannian surfaces with one Killing vector field}
A general approach for constructing Riemannian surfaces with one Killing vector
field which are maximally extended along geodesics is given in \cite{Katana09}.
Here we briefly describe the method and summarize the rules.

Suppose we have a Lorentzian signature metric
\begin{equation*}
  ds^2=\Phi(q)(d\tau^2-d\s^2).
\end{equation*}
To go to the positive or negative definite metric, as usual in physical
literature, we perform complex rotation of the time coordinate
$\tau\mapsto i\rho$. Then we get the Riemannian metric
\begin{equation}                                                  \label{eemtvi}
  ds^2=-\Phi(q)(d\s^2+d\rho^2).
\end{equation}
The sign of the conformal factor $\Phi(q)$ is not fixed, and we consider both
positive and negative definite metrics. The conformal factor is assumed to
depend on one independent argument $q$ related to coordinate $\s$ by ordinary
differential equation
\begin{equation}                                                  \label{emdmee}
  \left|\frac{dq}{d\s}\right|=\left|\Phi(q)\right|.
\end{equation}
Metric (\ref{eemtvi}) is precisely the metric (\ref{ebogae}), (\ref{ezintk})
obtained in the previous section up to inessential total sign. It has at least
one Killing vector field $\pl_\rho$, its length being equal to
$(\pl_\rho,\pl_\rho)=-\Phi(q)$.

We admit that the conformal factor $\Phi(q)$ has zeroes and singularities in the
finite set of points $q_i$, $i=1,\dots,k$ including infinities $q_1=-\infty$ and
$q_k=\infty$. Thus, the conformal factor has definite sign on every interval
$(q_i,q_{i+1})$. We assume power behaviour of the conformal factor near each
boundary point:
\begin{align}                                                     \label{ecfapo}
  |q_i|<\infty:&\qquad \Phi(q)\sim(q-q_i)^m,\qquad m\ne 0,
\\                                                                \label{ecfasi}
  |q_i|=\infty:&\qquad \Phi(q)\sim q^m.
\end{align}

In two dimensions, singularities of the curvature tensor are determined by
singularities of the scalar curvature
\begin{equation}                                                  \label{eriecl}
  R=-\Phi''.
\end{equation}
That is the curvature singularities occur at
\begin{align}                                                     \label{esfpsc}
  |q_i|<\infty:&\qquad m<0,\quad 0<m<1,\quad 1<m<2,
\\                                                                \label{esfpcc}
  |q_i|=\infty:&\qquad m>2.
\end{align}
It means that the surface cannot be extended along geodesics through these
points.

The domain of metric (\ref{eemtvi}) on the $\s,\rho$ plain depends on the form
of the conformal factor. It is clear that $\rho\in\MR$. For every interval
$q\in(q_i,q_{i+1})$, there is finite, semi-infinite, or infinite interval of
coordinate $\s$ depending on convergence or divergence of the integral
\begin{equation}                                                  \label{eferin}
  \s_{i,i+1}\sim\int^{q_i,q_{i+1}}\!\!\!\frac{dq}{\Phi(q)}
\end{equation}
at the boundary points:
\begin{equation}                                                  \label{eboucl}
\begin{array}{rl}
  |q_i|<\infty:\qquad &\left\lbrace
  \begin{array}{rll}
  m<1,  \quad &\text{converge,}  \\
  m\ge1,\quad &\text{diverge,}
  \end{array}\right. \\
  |q_i|=\infty:\qquad &\left\lbrace
  \begin{array}{rll}
  m\le1,\quad &\text{diverge,}\\
  m>1,  \quad &\text{converge,}
  \end{array}\right.
\end{array}
\end{equation}
For example, if on both sides of the interval $(q_i,q_{i+1})$ the integral
diverge, then $\s\in(-\infty,\infty)$ and the metric is defined on the whole
plane, $\s,\rho\in\MR^2$.

In the Lorentzian case, there is the conformal block for each interval
$(q_i,q_{i+1})$ which are glued together along horizons \cite{Katana00A}.
We shall see later that the ``light-like'' geodesics with the asymptotic
$\rho=\pm\s+\const$ are absent for the Riemannian signature metric, and there is
no need for continuation of the solution across the points $q_i$.

To construct global surfaces, we have to analyze behaviour of geodesics. They
can be analyzed analytically due to the existence of the Killing vector field.
The following statement is proved in \cite{Katana09}.
\begin{theorem}                                                   \label{textre}
Every geodesic for $\Phi>0$ and $dq/d\s>0$ belongs to one of the following
classes.
\newline
1. Straight geodesics
\begin{equation}                                                  \label{extlie}
  \rho=\pm\s+\const,
\end{equation}
exist only for the Euclidean metric $\Phi=\const$, the canonical parameter can
be chosen in the form $\s=t$.\\
2. The form of general type geodesics is defined by the equation
\begin{equation}                                                  \label{extgee}
  \frac{d\rho}{d\s}=\pm\frac1{\sqrt{-1+C\Phi}},
\end{equation}
where $C>0$ is an integration constant. The corresponding canonical parameter
is given by any of two equations:
\begin{align}                                                     \label{exttpe}
  \dot\s  &=\pm\frac{\sqrt{-1+C\Phi}}\Phi,
\\                                                                \label{extgpe}
  \dot\rho&=\frac1\Phi.
\end{align}
In addition, the signs in Eqs.~(\ref{extgee}) and (\ref{exttpe}) must be chosen
simultaneously.\newline
3. Straight geodesics parallel to the $\s$ axis go through every point
$\rho=\const$. The canonical parameter is defined by
\begin{equation}                                                  \label{extste}
  \dot\s=\frac1{\sqrt \Phi}.
\end{equation}
4. Straight degenerate geodesics parallel to the $\rho$ axis and going through
the points $\s_0=\const$, where the equation holds
\begin{equation}                                                  \label{extdce}
  \Phi'(\s_0)=0.
\end{equation}
The corresponding canonical parameter can be chosen as
\begin{equation}                                                  \label{extdge}
  t=\rho.
\end{equation}
\end{theorem}
Similar theorems are valid for all other signs of $\Phi$ and $dq/d\s$.

The important note is that Eqs.~(\ref{extgee}) and (\ref{exttpe}) differ from
the Lorentzian case by the
sign of unity under the square root. As the consequence, there are no general
type geodesics with the light-like asymptotics $\rho=\pm\s+\const$, and we do
not have to continue the surface across boundary points $q_i$ corresponding to
horizons.

The qualitative analysis of geodesics is given in \cite{Katana09}. It shows that
incompleteness of geodesics on the strip $\s\in(\s_i,\s_{i+1})$,
$\rho\in(-\infty,\infty)$ is entirely defined by the behavior of straight
geodesics parallel to the $\s$ axis at points $\s_i,\s_{i+1}$. In its turn,
their completeness is given by the convergence of the integral
\begin{equation}                                                  \label{extgpt}
  \underset{q\to q_{i+1}}\lim t\to\int^{q_{i+1}}\!\!\!\frac{dq}{\sqrt{|\Phi}|}.
\end{equation}
These geodesics are incomplete at finite points $|q_i|<\infty$ for $m<2$. At
infinite points $|q_i|=\infty$, they are incomplete for $m>2$. Geodesics are
complete in all other cases. Taking into account that continuation of the
surface can be made only through the points of regular curvature tensor, we see
that continuation must be performed only through the simple zero, $m=1$, at a
finite point $|q_i|<\infty$.

The following procedure for continuation of Riemannian surfaces with metric
(\ref{eemtvi}) is also useful for visualization of these surfaces.
We identify points $\rho$ and $\rho+L$, where $L>0$ is an arbitrary constant,
making a cylinder from the plane $(\s,\rho)\in\MR^2$. This is always possible
because metric components do not depend on $\rho$. The length $P$ of the
directing circle is defined by the conformal factor. Up to a sign, it can take
any value
\begin{equation}                                                  \label{elenci}
  P^2=-L^2\Phi(q)\overset{q\to\pm\infty}\longrightarrow
  \begin{cases} 0, & \Phi(\pm\infty)=0,
  \\ \const\ne0, & \Phi(\pm\infty)=\const\ne0,
  \\ \mp\infty & \Phi(\pm\infty)=\pm\infty, \end{cases}
\end{equation}
at infinite boundaries. At finite points $q\to q_i\ne\pm\infty$, the length can
take only two values:
\begin{equation}                                                  \label{elencu}
  P^2=-L^2\Phi(q)\overset{q\to q_i}\longrightarrow
  \begin{cases} 0, & \Phi(q_i)=0,
  \\ \mp\infty & \Phi(q_i)=\pm\infty. \end{cases}
\end{equation}
The plane $(\s,\rho)\in\MR^2$ is the universal covering of the cylinder.

Properties of boundary points $q_i$ are summarized in Table \ref{tprobo}.
\begin{table}[hbt]
\begin{center}
\begin{tabular}{|c|c|c|c|c|c|c|}
     \multicolumn{7}{c}{$|q_i|<\infty$}\\ \hline
     & $m<0$ & $0<m<1$ & $m=1$ & $1<m<2$ & $m=2$ & $m>2$ \\ \hline
$R$    & $\infty$ & $\infty$ & $\const$ & $\infty$ & $\const$ & $0$ \\
$\s_i$ & $\const$ & $\const$ & $\infty$ & $\infty$ & $\infty$ & $\infty$ \\
$P^2$    & $\infty$ & $0$ & $0$ & $0$ & $0$ & $0$ \\
Completeness& $-$ & $-$ & $-$ & $-$ & $+$ & $+$ \\  \hline
     \multicolumn{7}{c}{$|q_i|=\infty$}\\ \hline
     & $m<0$ & $m=0$ & $0<m\le 1$ & $1<m<2$ & $m=2$ & $m>2$ \\ \hline
$R$    & $0$ & $0$ & $0$ & $0$ & $\const$ & $\infty$ \\
$\s_i$ & $\infty$ & $\infty$ & $\infty$ & $\const$ & $\const$ & $\const$ \\
$P^2$    & $0$ & $\const$ & $\infty$ & $\infty$ & $\infty$ & $\infty$ \\
Completeness& $+$ & $+$ & $+$ & $+$ & $+$ & $-$ \\ \hline
\end{tabular}
\end{center}
 \caption{Properties of boundary points depending on the exponent $m$. Symbol
 $\const$ denotes nonzero constant in the rows of scalar curvature and lengths
 of directing circle up to a sign.
 \label{tprobo}}
\end{table}

Figure~\ref{fsurec} shows the form of Riemannian surfaces near boundary
points $q_{i+1}$ after identification $\rho\sim\rho+L$. The surfaces near points
$q_i$ have the same form but opposite direction. The surface corresponding to
an interval $(q_i,q_{i+1})$ is obtained by gluing two such surfaces for boundary
points $q_i$ and $q_{i+1}$.
\begin{figure}[tbp]
\hfill\includegraphics[width=.95\textwidth]{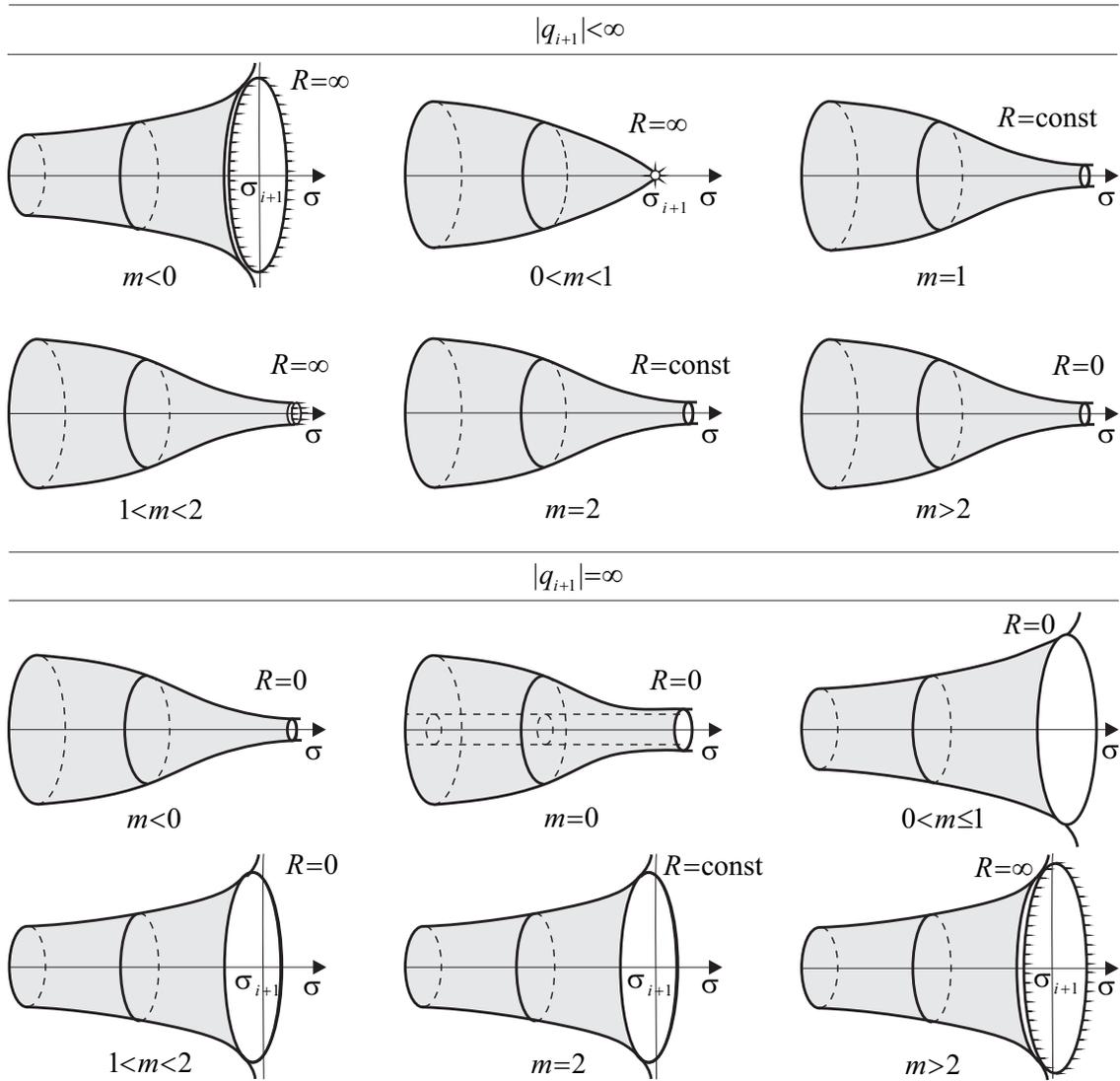}
\hfill {}
\\
\centering
 \caption{The surface near the boundary point $q_{i+1}$ after identification
 $\rho\sim\rho+L$. We assume that the coordinate $\s$ increases from the
 left to the right and $\s_{i+1}>\s_i$.
 \label{fsurec}}
\end{figure}
The table \ref{tprobo} shows that the point $q_i$ is geodesically incomplete
and has regular curvature only at finite points $|q_i|<\infty$ for $m=1$.

The continuation of geodesics is performed as follows. The corresponding
boundary $q_i$ is really a point because $P^2(q_i)=0$. Moreover this infinite
point $\s=\infty$ in the plane $\rho,\s$ lies, in fact, at finite distance
because geodesics reach it at a finite value of the canonical parameter.
Performing the coordinate transformation \cite{Katana09} one can easily show
that this is the conical singularity with the deficit angle
\begin{equation}                                                  \label{edefae}
  2\pi\theta:=\frac{L|\Phi'_i|}2-2\pi,\qquad \Phi'_i:=\Phi'|_{q=q_i}.
\end{equation}
For the particular value
\begin{equation}                                                  \label{qvbttf}
  L=4\pi/|\Phi'_i|,
\end{equation}
the conical singularity is absent.

Thus, continuation across the point $|q_i|<\infty$ for $m=1$ has no meaning
in general because this is a conical singularity. In the particular case
(\ref{qvbttf}), geodesics are continued as shown in Fig.~\ref{fcorec}. The
straight geodesics parallel to the axis $\s$ and going through the points $\rho$
and $\rho+L/2$ are two halves of one geodesic as shown in the figure.
\begin{figure}[htb]
\hfill\includegraphics[width=.35\textwidth]{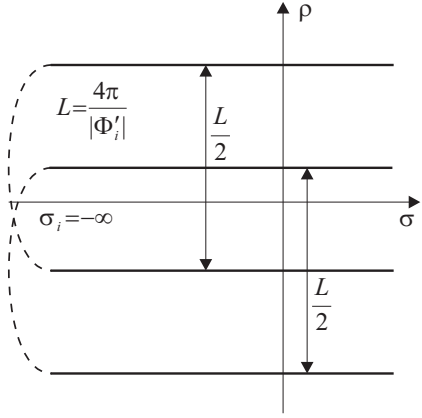}
\hfill {}
\caption{Continuation of straight geodesics going through points $\rho$ and
$\rho+L/2$ in the absence of conical singularity $L=4\pi/|\Phi'_i|$.
The identification occur at the point $\s_i=-\infty$.}
 \label{fcorec}
\end{figure}
In the absence of conical singularity, the fundamental group in trivial, and the
corresponding Riemannian surface is the universal covering itself.

\begin{figure}[htb]
\hfill\includegraphics[width=.95\textwidth]{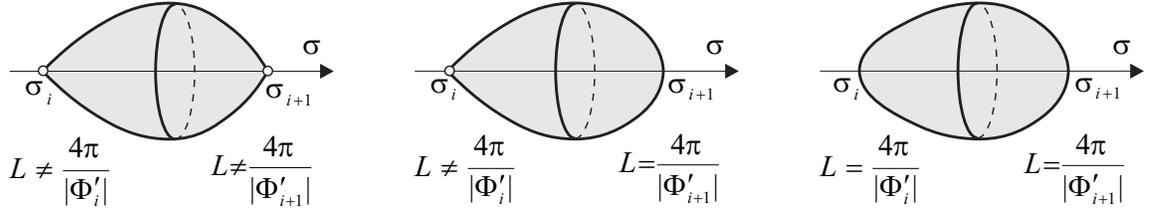}
\hfill {}
\\
\centering
 \caption{Three Riemannian surfaces for the interval $(q_i,q_{i+1})$, when the
 conformal factor has asymptotics $\Phi\sim q-q_{i,i+1}$ at boundary points.
 \label{fglmon}}
\end{figure}

If the conformal factor has asymptotics $\Phi\sim(q-q_i)$ and
$\Phi\sim(q_{i+1}-q)$ on both sides of the interval $(q_i,q_{i+1})$, then the
surface must be continued on both points $\s=\pm\infty$. After the
identification we get zero, one, or two conical singularities as shown in
Fig.~\ref{fglmon}. These Riemannian surfaces are topologically a sphere with,
possibly, one or two conical singularities.

To summarize, we formulate the rules for construction of maximally extended
Riemannian surfaces with metric (\ref{eemtvi}).
\begin{enumerate}
\item After identification $\rho\sim\rho+L$, there is unique maximally extended
Riemannian surface corresponding to every interval $(q_i,q_{i+1})$ which is
obtained by gluing together two surfaces depicted in Fig.~\ref{fsurec} for two
boundary points $q_i$ and $q_{i+1}$.
\item In all cases, except the absence of conical singularity $|q_i|<\infty$,
$L|\Phi'_i|\ne4\pi$ or $|q_{i+1}|<\infty$, $L|\Phi'_{i+1}|\ne4\pi$, the strip
$\s\in(\s_i,\s_{i+1})$, $\rho\in\MR$ with metric (\ref{eemtvi}) is the universal
covering space for the corresponding maximally extended surface.
\item In the absence of one of conical singularities, $|q_i|<\infty$,
$L|\Phi'_i|=4\pi$, or $|q_{i+1}|<\infty$, $L|\Phi'_{i+1}|=4\pi$, the surface
obtained by the identification $\rho\sim\rho+L$ is the universal covering space
itself.
\end{enumerate}
\subsection{Lorentz-invariant and planar solutions for $K^{(g)}=1$
                                                                 \label{solhyp}}
Let us now classify all global solutions in the case {\sf C}.
We note first that the cases $K^{(g)}=1$ and $K^{(g)}=-1$ are related by
insignificant permutation of the first two coordinates $x^0\leftrightarrow x^1$
and therefore are equivalent. We choose $K^{(g)}=1$. The metric for one-sheeted
hyperboloid in hyperbolic polar coordinate system $\theta$, $\vf$ is
\begin{equation}                                                  \label{qingtf}
  ds^2=d\theta^2-\cosh^2\theta d\vf^2.
\end{equation}
Hence, the four-dimensional metric of the space-time in Schwarzschild-like
coordinates has the form
\begin{equation}                                                 \label{egsmts}
  ds^2=q^2(d\theta^2-\cosh^2\theta d\vf^2)+\frac{dq^2}{\Phi(q)}+\Phi (q)d\rho^2,
\end{equation}
where the conformal factor
\begin{equation}                                                  \label{qcodlj}
  \Phi=1-\frac{2M}q+\frac{Q^2}{q^2}+\frac{\Lm q^2}3,
\end{equation}
has the same form as in the spherically symmetric case up to a sign of the
cosmological constant $\Lm$.

The metric on surface $\MV$ can be negative ($\Phi<0$) or positive
($\Phi>0$) definite depending on the constants $M$, $Q$, $\Lm$ and the interval
$(q_i,q_{i+1})$. For negative definite metric on $\MV$ the signature of the
space-time is $(+---)$, and the coordinate $\theta$ plays the role of time.
Therefore the time-like coordinate takes values on the whole real line
$\theta\in\MR$, and three-dimensional spatial sections $\theta=\const$ are given
by the product of the circle $\vf\in[0,2\pi]$ on the Riemannian surfaces $\MV$
which will be constructed later. If $\MU$ is the universal covering space of
one-sheeted hyperboloid, then the three-dimensional space is the product
$\MR\times\MV$. If the surface $\MV$ has singularity, then it corresponds to
singular timelike surface in the space-time.

For the positive definite metric on $\MV$, the signature of the four-dimensional
metric is $(+-++)$, and the angle $\vf$ is the time-like coordinate. It takes
values on a circle $\vf\in[0,2\pi]$ for one-sheeted hyperboloid $\MU=\ML^2$, and
three-dimensional spatial sections are the product of real line $\theta\in\MR$
on surface $\MV$. The corresponding space-time contains closed time-like curves
(including geodesics), if we did not choose the universal covering space for the
one-sheeted hyperboloid.

The unexpected is the following observation. Assume that metric (\ref{egsmts})
is given and the values of $M$, $Q$, and $\Lm$ are fixed. Then, in general,
there are  several global topologically disconnected solutions of different
signatures for the same set of $M$, $Q$, and $\Lm$. Indeed, the signature of
metric (\ref{egsmts}) is defined by the sign of the conformal factor which may
have different signs on different intervals $(q_i,q_{i+1})$.

Now we classify maximally extended surfaces $\MV$ in the case {\sf C} using the
method reviewed in the previous section (more details are contained in
\cite{Katana09}). Let us remind the reader that we
consider nonnegative $M\ge0$ and positive $Q>0$. We start with the simplest case
when the conformal factor can be analyzed quantitatively.

\subsubsection{Zero cosmological constant $\Lm=0$.}
\begin{figure}[t]
\hfill\includegraphics[width=.8\textwidth]{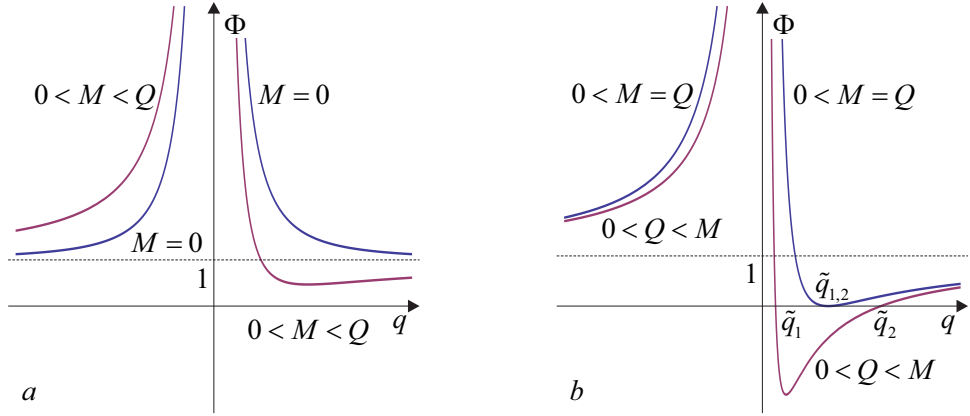}
\hfill {}
\centering\caption{The conformal factor for ({\it a}) $M=0$,
$Q>0$ and $0<M<Q$; ({\it b}) $0<M=Q$ и $0<Q<M$.}
\label{f31ConfFacEM+G}
\end{figure}

\begin{figure}[hbt]
\hfill\includegraphics[width=.8\textwidth]{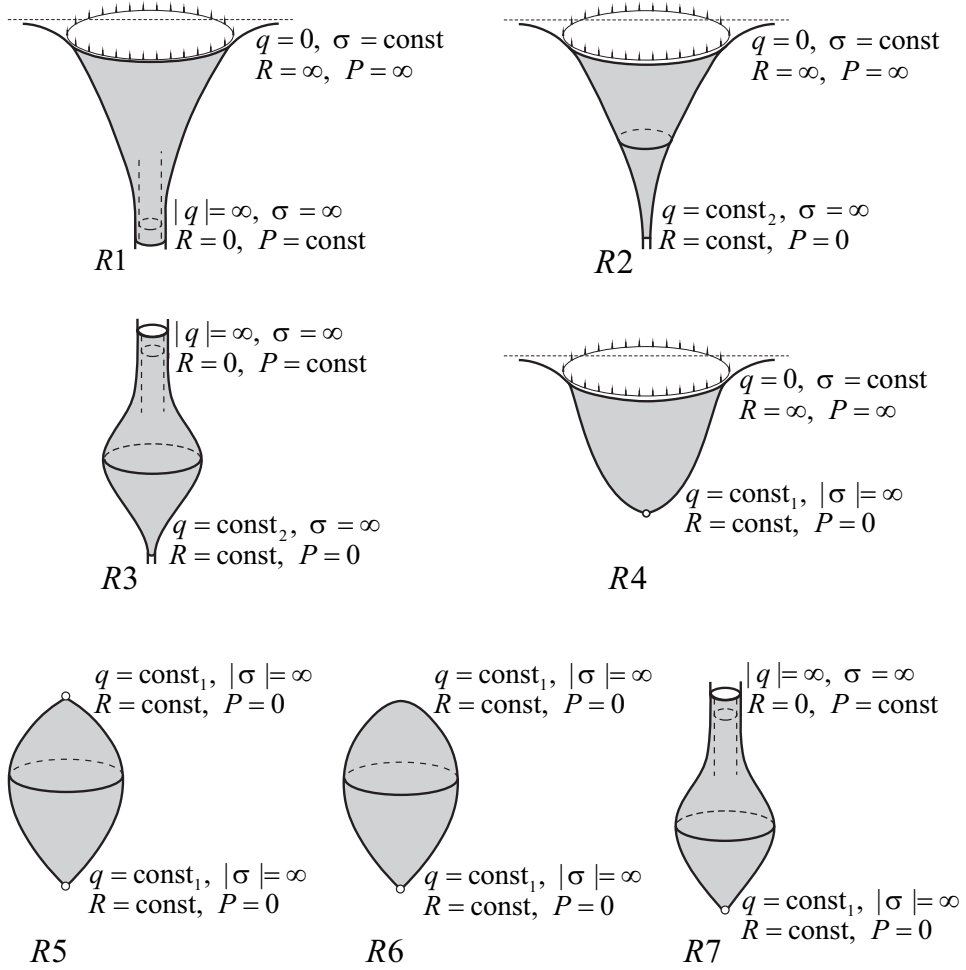}
\hfill {}
\centering\caption{Riemannian surfaces $\MV$ for $K^{(g)}=1$ and $\Lm=0$.
The inscription $q=\const_{1,2}$ means that the conformal factor $\Phi(q)$ has
zero of the first or second order at this point.}
\label{f31EuclSurEM+G}
\end{figure}

For $\Lm=0$, the conformal factor (\ref{unnvbf}) is
\begin{equation}                                                  \label{ubvxfr}
  \Phi=1-\frac{2M}q+\frac{Q^2}{q^2},\qquad M\ge0,\quad Q>0,\qquad q\ne0.
\end{equation}
If $M=0$ and $Q>0$ or $0<M<Q$, then the conformal factor does not have zeroes
and is shown in Fig.~\ref{f31ConfFacEM+G}, {\it a}. This case corresponds to the
naked singularity for the Lorentzian signature metric.
We have two global solutions
for fixed $M$ and $Q$ corresponding to intervals $(-\infty,0)$ and $(0,\infty)$.
They have the form $\MM=\ML^2\times\MV$, where $\ML^2$ is the universal covering
of one-sheeted hyperboloid. The surfaces $\MV$ for intervals $(-\infty,0)$ and
$(0,\infty)$ coincide topologically and differ only by the orientation:
$\s\to-\s$. The existence of local minimum in the conformal factor for $0<M<Q$
leads to the appearance of degenerate and oscillating geodesics without topology
change of surface $\MV$. The corresponding surface $\MV$ is topologically a half
infinite cylinder denoted by $R1_{(++)}$ and shown in Fig.~\ref{f31EuclSurEM+G}.
Indices $(++)$ denote the metric signature on the surface $\MV$.
$\sign h_{\mu\nu}=(++)$ because the conformal factor is positive. Consequently,
the signature of the four-dimensional metric is $(-+++)$. This surface is
geodesically incomplete for finite value $\s_0$ corresponding to $q=0$.
The two-dimensional scalar curvature is singular there, and the surface cannot
be extended. In addition, the length of the circle $\s=\const$ goes to infinity
when $\s\to\s_0$. At $q\to\pm\infty$ the surface is geodesically complete, and
$\s_{\pm\infty}=\pm\infty$. There, the curvature tends to zero and the length of
the circle $\s=\const$ goes to some finite value. Spatial sections $x^0=\const$
have the form $\MS^1\times\MV$, if $\MU$ is the one-sheeted hyperboloid, or
$\MR\times\MV$, if $\MU$ is the universal covering of one-sheeted hyperboloid.
From physical standpoint, these solutions describe infinite evolution of the
domain wall of curvature singularity located at $\s_0$.

The parameter $q$ and coordinate $\s$ change in the vertical direction in
Fig.~\ref{f31EuclSurEM+G}. For some surfaces, the coordinate $\s$ increases
upwards and for others increases downwards. We do not distinguish these cases
because topologically they are equivalent.

For $0<M=Q$ the conformal factor has one zero of the second order
\begin{equation}                                                  \label{unbgtl}
  \Phi=\frac{(q-M)^2}{q^2},
\end{equation}
located at the point $q_{1,2}=M$. It corresponds to the extremal black hole in
the Lorentzian case. There is the same surface
$\MV=R1_{(++)}$ corresponding to the interval $q\in(-\infty,0)$ as in the
previous case. For positive $q$, we have two global solutions relevant to
the intervals $(0,q_{1,2})$ and $(q_{1,2},\infty)$.
For $q\in(0,q_{1,2})$, the surface is $\MU=R2_{(++)}$ and shown in
Fig.~\ref{f31EuclSurEM+G}. This surface is geodesically incomplete at $q=0$
where two-dimensional scalar curvature becomes infinite. Thus this solution
describes infinite evolution of the domain wall of curvature singularity but, in
contrast to the previous case, the length of the circle $\s=\const$ vanishes as
$q\to q_{1,2}$.

If $q\in(q_{1,2},\infty)$, then the surface has the form $R3_{(++)}$ in
Fig.~\ref{f31EuclSurEM+G}. This surface is topologically an infinite cylinder,
does not have singularities, and geodesically complete.

If the inequalities $0<Q<M$ hold, then the conformal factor has two positive
simple zeroes:
\begin{equation}                                                  \label{uvvbxb}
  \Phi=\frac{(q-q_1)(q-q_2)}{q^2},
\end{equation}
where
\begin{equation*}
  q_{1,2}=M\pm\sqrt{M^2-Q^2}.
\end{equation*}
It corresponds to the Reissner--Nordstr\"om black hole.
We have four global solutions in this case corresponding to the intervals:
$(-\infty,0)$, $(0,q_1)$, $(q_1,q_2)$, and $(q_2,\infty)$. For
$q\in(-\infty,0)$ the surface is the same as before: $\MV=R1_{(++)}$.

If $q\in(0,q_1)$, then the surface has the form $\MV=R4_{(++)}$. The
hollow circle denotes possible conical singularity at $q=q_1$, its
existence depending on the identification along coordinate $\rho$. This surface
is geodesically incomplete at the point $q=0$, where the curvature has
singularity. If the conical singularity does exist, then, from physical point of
view, this solution describes infinite evolution of the cosmic string surrounded
by the domain wall of curvature singularity.

For $q\in(q_1,q_2)$ the metric on surface $\MV$ is negative
definite because the conformal factor is negative. Depending on the
identification along coordinate $\rho$, this surface may have one or two
conical singularities. There are two possible surfaces:
$\MV=R5_{(--)}$ (two conical singularities at points $q_1$ and $q_2$), or
$\MV=R6_{(--)}$ (one conical singularity at $q_1$ or at $q_2$. It is not
possible to eliminate simultaneously both conical singularities. In these cases,
global solutions describe infinite evolution of one or two cosmic strings. If
the surface $\MU=\ML^2$ is the one-sheeted hyperboloid, then spatial sections
are compact, and cosmic strings have the form of a circle. If geodesics are
symmetrically continued through the conical singularities, then surfaces
$\MV=R5_{(--)}$ and $\MV=R6_{(--)}$ are geodesically complete.

For $q\in(q_2,\infty)$, the surface has the form $\MV=R7_{(++)}$. In this
case, the space-time is geodesically complete and describes infinite evolution
of, possibly, one cosmic string at point $q_2$.
\subsubsection{Nonzero cosmological constant $\Lm\ne0$}
The analysis of the conformal factor with nonzero cosmological constant can be
performed qualitatively. The form of Riemannian surface $\MV$ is defined by
zeroes of the conformal factor (\ref{qcodlj}) which we rewrite in the form
\begin{equation}                                                  \label{unnvbf}
  \Phi(q)=1-\frac{2M}q+\frac{Q^2}{q^2}+\frac{\Lm q^2}3
  =\frac{\psi(q)+3Q^2}{3q^2},
\end{equation}
where
\begin{equation}                                                  \label{ubbcgj}
  \psi(q):=\Lm q^4+3q^2-6Mq
\end{equation}
is the auxiliary function needed for the following analysis.

When $Q\ne0$, the conformal factor (\ref{unnvbf}) has the pole of the second
order at $q=0$. Zeroes of the conformal factor coincide with zeroes of the
function $\psi(q)+3Q^2=0$. To find the number and type of zeroes of this
function we analyse qualitatively the function $\psi(q)$ and then move it
upwards by $3Q^2$.

Differentiate twice the auxiliary function (\ref{ubbcgj}):
\begin{equation}                                                  \label{unmjqd}
\begin{split}
  \psi'(q)=&4\Lm q^3+6q-6M,
\\
  \psi''(q)=&12\Lm q^2+6=6(2\Lm q^2+1).
\end{split}
\end{equation}

It is easy to find asymptotics of the function $\psi(q)$ ($\Lm\ne0$) and its
derivatives at $q=0$ and $q\to\infty$:
\begin{align}                                                          \nonumber
  \psi(0)=&~~0, & \psi(q\to\infty)\approx&\Lm q^4,
\\                                                                \label{edswhy}
  \psi'(0)=&-6M, & \psi'(q\to\infty)\approx&4\Lm q^3,
\\                                                                     \nonumber
  \psi''(0)=&~~6, & \psi''(q\to\infty)\approx&12\Lm q^2.
\end{align}

The points of inflection are defined by equality $\psi''=0$, which implies two
inflection points $q^*=\pm1/\sqrt{-2\Lm}$ for nonnegative cosmological constant
and arbitrary $M$. At these points
\begin{equation}                                                  \label{uvbzfd}
  \psi(q^*)=-\frac5{4\Lm}\mp\frac{6M}{\sqrt{-2\Lm}},\qquad\Lm<0.
\end{equation}
The case, when the first derivative is equal to zero,
\begin{equation}                                                  \label{ejhytg}
  \psi'(q^*)=\pm\frac4{\sqrt{-2\Lm}}-6M=0\qquad\Rightarrow\qquad
  M=\pm\frac13\sqrt{-\frac2\Lm}
\end{equation}
at the inflection point, is of particular interest for global solutions.
For positive $M$, which we only consider, $q^*=1/\sqrt{-2\Lm}>0$. After the
shift by $3Q^2:=-\psi(q^*)$, the conformal factor has the third order zero.
More precisely, for
\begin{equation*}
  q^*=\frac1{\sqrt{-2\lm}},\qquad M=\frac13\sqrt{-\frac2\Lm},\qquad
  Q^2=-\frac1{4\Lm},\qquad\Lm<0
\end{equation*}
the equality holds
\begin{equation}                                                  \label{ubdngr}
  \psi(q)+3Q^2=\Lm(q-q^*)^3\left(q+\frac3{\sqrt{-2\Lm}}\right).
\end{equation}

Finding of zeroes of the function $\psi(q)+3Q^2$ is more complicated. This
polynomial of the fourth order can have not more then four real roots
depending on the values of $\Lm\ne0$, $M\ge0$, and $Q>0$. To determine the
zeroes type, we need to know local extrema of functions $\psi(q)$, which become
zeroes of the second or third order after shifting on corresponding valued of
$3Q^2$.

Local extrema of function $\psi$ are defined by the cubic equation (the solution
is given, e.g.\ in \cite{KorKor68})
\begin{equation}                                                  \label{uvskju}
  q^3+\frac3{2\Lm} q-\frac{3M}{2\Lm}=0.
\end{equation}
There are three possible cases depending on the value of the constant
\begin{equation}                                                  \label{ebgtdj}
  \Upsilon:=\frac1{8\Lm^3}+\frac{9M^2}{16\Lm^2}.
\end{equation}
\subsubsection{Negative cosmological constant $\Lm<0$}
For $\Lm<0$, there are three cases for Eq.~(\ref{uvskju}):
\begin{align*}
  \Upsilon>0\quad\Leftrightarrow\quad M>&\frac13\sqrt{-\frac2\Lm} \qquad
  \text
  {--\quad \parbox[c]{0.5\textwidth}{one real and \\
  two complex conjugate roors,}}
\\[6pt]
  \Upsilon=0\quad\Leftrightarrow\quad M=&\frac13\sqrt{-\frac2\Lm} \qquad
  \text{--\quad \parbox[c]{0.5\textwidth}{three real roots \\
  (at least two of them coinside),}}
\\[6pt]
  \Upsilon<0\quad\Leftrightarrow\quad M<&\frac13\sqrt{-\frac2\Lm} \qquad
  \text{--\quad three real different roots.}
\end{align*}

We start with the simplest case $\Upsilon=0$, when two roots coincide. This
equality implies the restriction on the ``mass'':
\begin{equation}                                                  \label{edbgdt}
  \Upsilon=0\qquad\Leftrightarrow\qquad M=\frac13\sqrt{-\frac2\Lm}.
\end{equation}
The roots of Eq.~(\ref{uvskju}) have the simple form
\begin{equation}                                                  \label{unbhsk}
  \qquad q_1=-\sqrt{-\frac2\Lm}, \qquad
  q_{2,3}=\frac12\sqrt{-\frac2\Lm},
\end{equation}
There are one simple negative root and one positive root of the second order for
positive ``mass'' (\ref{edbgdt}). We shall need the values of the auxiliary
function at $q_{2,3}$:
\begin{equation}                                                  \label{ubbans}
  \psi_{2,3}:=\psi(q_{2,3})=\frac3{4\Lm},
\end{equation}
at which the conformal factor has zero of the second order.

If the inequality $\Upsilon<0$ holds, then real roots of the cubic equation
(\ref{uvskju}) are different and equal to (see e.g.\ \cite{KorKor68})
\begin{equation}                                                  \label{ubcndt}
  q_3=\sqrt{-\frac2\Lm}\cos\frac\al3,\qquad
  q_{2,1}=-\sqrt{-\frac2\Lm}\cos\left(\frac\al3\pm\frac\pi3\right),
\end{equation}
where
\begin{equation*}
  \cos\al:=-3M\sqrt{-\frac\Lm2}.
\end{equation*}
The angle $\al\in\big[\frac\pi2,\frac{3\pi}2\big]$ because we consider only
nonnegative $M$. It implies the existence of one negative root $q_1$ and two
positive roots $q_2$ and $q_3$. We enumerate the roots in Eq.~(\ref{ubcndt}) in
such a way that in the limit
\begin{equation*}
  M\to\frac13\sqrt{-\frac2\Lm}
\end{equation*}
the roots $q_{1,2,3}$ take values (\ref{unbhsk}).

If $\Upsilon>0$, then there is only one negative root $q_1$, the exact value of
which can be written but it does not matter.

Let us introduce the notation
\begin{equation*}
  \psi_1:=\psi(q_1),\qquad\psi_2:=\psi(q_2),\qquad\psi_3:=\psi(q_3).
\end{equation*}
It is easily seen that $\psi_1>0$ and $\psi_2<0$. The value $\psi_3$ can be
either negative or positive: it depends on the relation between $\Lm$ and $M$.
Indeed, two relations must hold at point $\psi_3=0$:
\begin{equation*}
\begin{split}
  \psi(q)=&\Lm q^4+3q^2-6Mq=0,
\\
  \psi'(q)=&4\Lm q^3+6q-6M=0.
\end{split}
\end{equation*}
For the fixed cosmological constant, we have the system of two equations on
$M$ and $q$, which has the unique solution for positive $M$
\begin{equation}                                                  \label{uxbsrf}
  M=\frac1{3\sqrt{-\Lm}},\qquad \bar q_3=\frac1{\sqrt{-\Lm}}.
\end{equation}
Thus, the following inequalities hold:
\begin{equation}                                                  \label{unsksj}
\begin{aligned}
  M>&\frac1{3\sqrt{-\Lm}}\qquad &\Rightarrow&\qquad & \psi_3<&0,
\\
  M=&\frac1{3\sqrt{-\Lm}} &\Rightarrow & & \psi_3=&0,
\\
  M<&\frac1{3\sqrt{-\Lm}} &\Rightarrow & & \psi_3>&0.
\end{aligned}
\end{equation}

\begin{figure}[hbt]
\hfill\includegraphics[width=.8\textwidth]{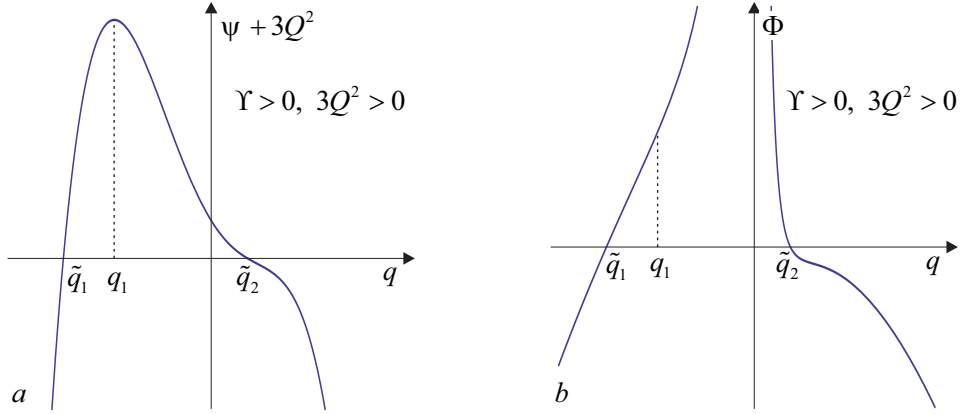}
\hfill {}
\centering\caption{Qualitative behavior of the function $\psi(q)+3Q^2$ and
conformal factor $\Phi(q)$ for $\Lm<0$, $\Upsilon>0$, and $3Q^2>0$.}
\label{f31psiPhi1GR+M}
\end{figure}

\begin{figure}[hbt]
\hfill\includegraphics[width=.8\textwidth]{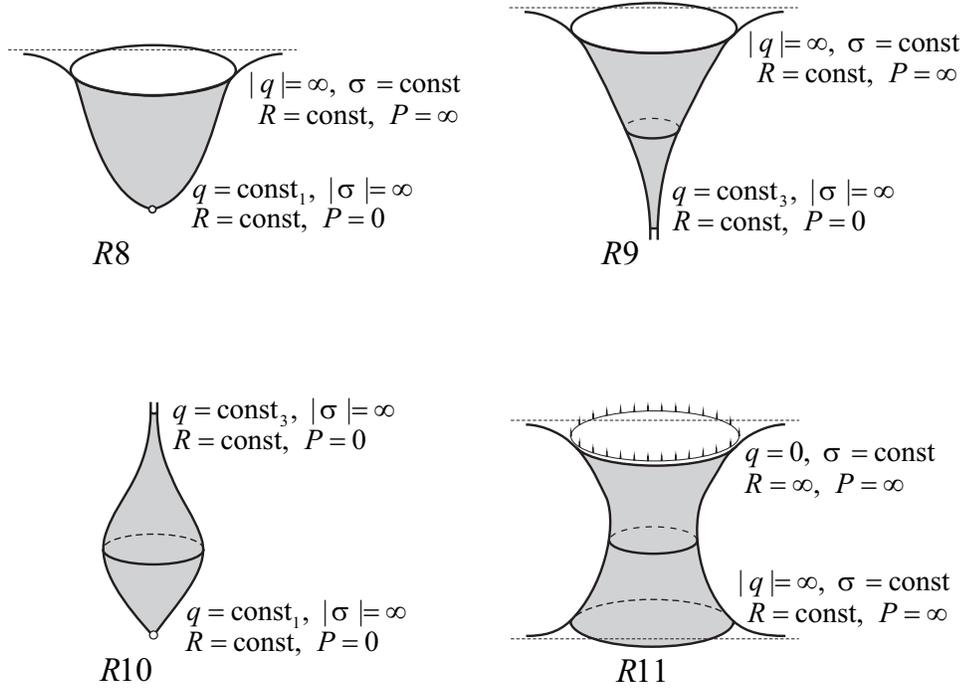}
\hfill {}
\centering\caption{Additional Riemannian surfaces $\MV$ for zero cosmological
constant $\Lm\ne0$.}
\label{f31RimSurLmGR+M}
\end{figure}

Let
\begin{equation}                                                  \label{usbdre}
\begin{aligned}
  &\Lm<0,\qquad & &\Upsilon>0,\qquad & &3Q^2>0,
\\
  &\Lm<0, & &\Upsilon=0, & &3Q^2\ne-\psi_{2,3},
\\
  &\Lm<0, & &\Upsilon<0, & &3Q^2>-\psi_2.
\end{aligned}
\end{equation}
Qualitative behavior of the function $\psi(q)+3Q^2$ and the conformal factor
$\Phi(q)$ for the upper row in the inequalities are shown in
Fig.~\ref{f31psiPhi1GR+M}. Zeroes
of the conformal factor are simple and located at points $\tilde q_1$ and
$\tilde q_2$. The qualitative behavior of the conformal factor for the second
and third rows of inequalities (\ref{usbdre}) is the same. Thus, in these cases,
the function $\psi(q)+3Q^2$ and consequently the conformal factor $\Phi(q)$ have
two simple roots each. As the consequence, there are four global solutions
corresponding to the intervals $(-\infty,\tilde q_1)$, $(\tilde q_1,0)$,
$(0,\tilde q_2)$, and $(\tilde q_2,\infty)$ when the inequality (\ref{usbdre})
holds. Global solutions for intervals $(-\infty,\tilde q_1)$ and
$(\tilde q_2,\infty)$ coincide topologically differing only by the orientation
$\s\to-\s$. The surfaces $\MV$ in this case have the form $R8_{(--)}$ and are
shown in Fig.~\ref{f31RimSurLmGR+M}.
These surfaces are topologically a plane $\MR^2$ with possible conical
singularity. They are geodesically complete outside the conical singularity and
describe infinite evolution of the cosmic string.

For intervals $(\tilde q_1,0)$ and $(0,\tilde q_2)$, the surfaces have the
form $R4_{(++)}$ and are depicted in Fig.~\ref{f31EuclSurEM+G}.

\begin{figure}[hbt]
\hfill\includegraphics[width=.8\textwidth]{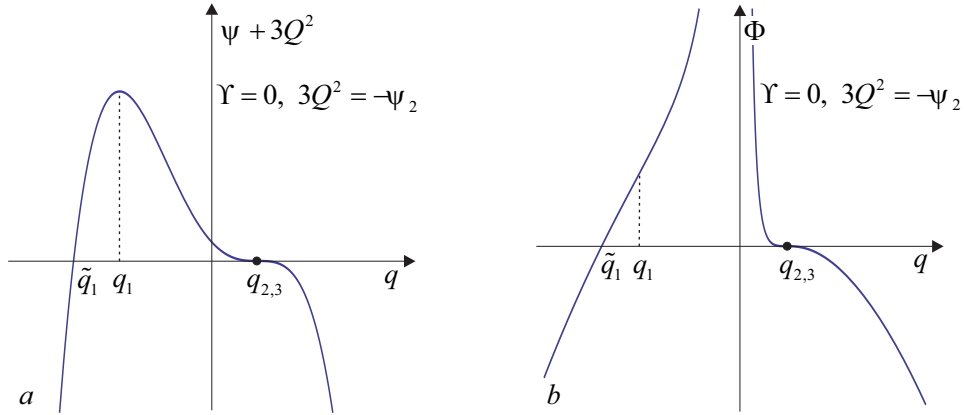}
\hfill {}
\centering\caption{Qualitative behavior of the function $\psi(q)+3Q^2$ and
conformal factor $\Phi(q)$ for $\Lm<0$, $\Upsilon=0$, and $3Q^2=-\psi_2$.}
\label{f31psiPhi2GR+M}
\end{figure}

In the case
\begin{equation}                                                  \label{unbchd}
  \Lm<0,\qquad\Upsilon=0,\qquad 3Q^2=-\psi_2,
\end{equation}
the conformal factor has two zeroes: the first zero $\tilde q_1$ is simple and
the second one $\tilde q_2=q_{2,3}$ is of the third order. The corresponding
function $\psi(q)+3Q^2$ and conformal factor $\Phi(q)$ are shown in
Fig.~\ref{f31psiPhi2GR+M}.
We have four global solutions corresponding to the intervals
$(-\infty,\tilde q_1)$, $(\tilde q_1,0)$, $(0,q_{2,3})$, and
$(q_{2,3},\infty)$ when equalities (\ref{unbchd}) hold. As before, surfaces
$R8_{(--)}$ and $R4_{(++)}$ correspond to the intervals $(-\infty,\tilde q_1)$
and $(\tilde q_1,0)$, respectively.

The global surface $\MV$ for the interval $(0,q_{2,3})$ has the form $R2_{(++)}$
in Fig.~\ref{f31EuclSurEM+G}. The scalar curvature vanish at the point $q_{2,3}$
because it is a zero of the third order.

New surface $R9_{(--)}$ in Fig.~\ref{f31RimSurLmGR+M} corresponds to the
interval $(q_{2,3},\infty)$. It is geodesically complete and does not contain
singularities.

Thus, we have constructed all global solutions for $\Lm<0$ and $\Upsilon\ge0$.
The case $\Upsilon<0$ includes more possibilities, because the auxiliary
function  $\psi(q)$ has from two to four zeroes. The case of two zeroes was
already described (the third row in Eq.~(\ref{usbdre})). Now we consider the
remaining cases.

\begin{figure}[hbt]
\hfill\includegraphics[width=.8\textwidth]{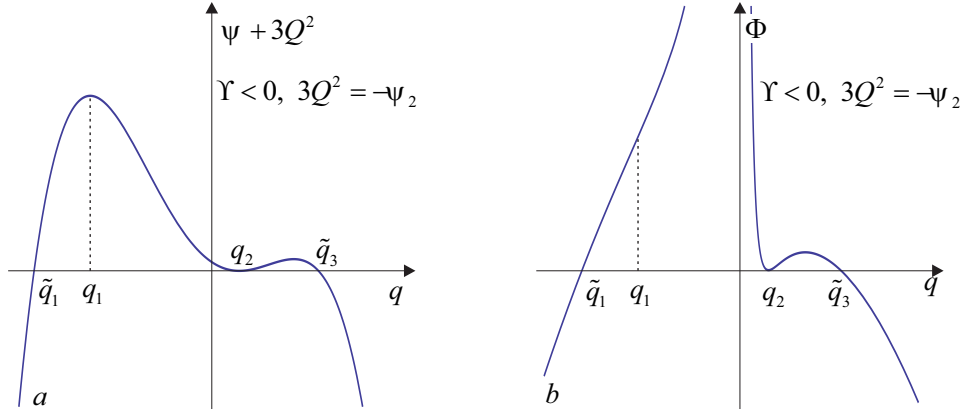}
\hfill {}
\centering\caption{Qualitative behavior of the auxiliary function $\psi(q)+3Q^2$
and conformal factor $\Phi(q)$ for $\Lm<0$, $\Upsilon<0$, and $3Q^2=-\psi_2$.}
\label{f31psiPhi3GR+M}
\end{figure}

Let
\begin{equation}                                                  \label{ubbdnh}
  \Lm<0,\qquad \Upsilon<0,\qquad 3Q^2=-\psi_2.
\end{equation}
Corresponding graphs of the auxiliary function and conformal factor are shown in
Fig.~\ref{f31psiPhi3GR+M}.

The conformal factor has three zeroes: two simple zeroes at points $\tilde q_1$,
$\tilde q_3$ and the zero of the second order at point $\tilde q_2=q_2$.
Therefore there are five global solutions in this case. Solutions for the
intervals $(-\infty,\tilde q_1)$, $(\tilde q_1,0)$, $(0,q_2)$, and
$(\tilde q_3,\infty)$ were already met: they are, respectively, $R8_{(--)}$,
$R4_{(++)}$, $R2_{(++)}$, and $R8_{(--)}$.

Global solution for the interval $(q_2,\tilde q_3)$ has the form $R10_{(++)}$ in
Fig.~\ref{f31psiPhi3GR+M}. It is topologically a plane $\MR^2$ with, possibly,
one conical singularity. If the conical singularity is present, then the
solution describes infinite evolution of the cosmic string. The unexpected
property of this solution is that the length of the circle located at geodesic
infinity $q\to q_2$ tends to zero, $P\to0$.

\begin{figure}[hbt]
\hfill\includegraphics[width=.8\textwidth]{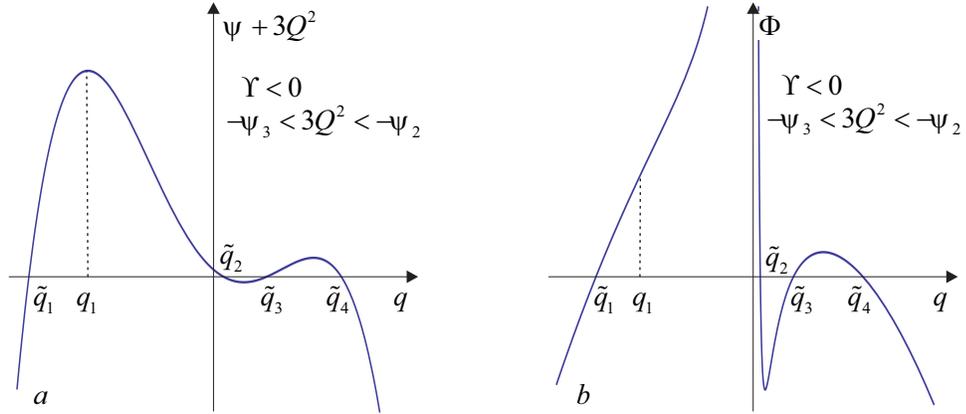}
\hfill {}
\centering\caption{Qualitative behavior of the function $\psi(q)+3Q^2$ and
conformal factor $\Phi(q)$ for $\Lm<0$, $\Upsilon<0$, and
$-\psi_3<3Q^2<-\psi_2$.}
\label{f33psiPhi4GR+M}
\end{figure}

If
\begin{equation}                                                  \label{ubcnvh}
  \Lm<0,\qquad\Upsilon<0,\qquad -\psi_3<3Q^2<-\psi_2,
\end{equation}
then the conformal factor has the maximal number of zeroes -- four. In this
case, the inequality $3Q^2>-\psi_3$ для $\psi_3\ge0$ holds automatically. The
graphs of the function $\psi(q)+3Q^2$ and conformal factor $\Phi(q)$ are given
in Fig.~\ref{f33psiPhi4GR+M}. All four zeroes of the conformal factor are
simple, and totally there are six global solutions which were already met:
\begin{equation}                                                  \label{ubdnsh}
\begin{aligned}
  (-\infty,\tilde q_1)&:~R8_{(--)}, & \qquad(\tilde q_2,\tilde q_3)&:~
  R5_{(--)},~R6_{(--)},
\\
  (\tilde q_1,0)&:~R4_{(++)}, & (\tilde q_3,\tilde q_4)&:~
  R5_{(++)},~R6_{(++)},
\\
  (0,\tilde q_2)&:~R4_{(++)}, & (\tilde q_4,\infty)&:~R8_{(--)}.
\end{aligned}
\end{equation}

\begin{figure}[hbt]
\hfill\includegraphics[width=.8\textwidth]{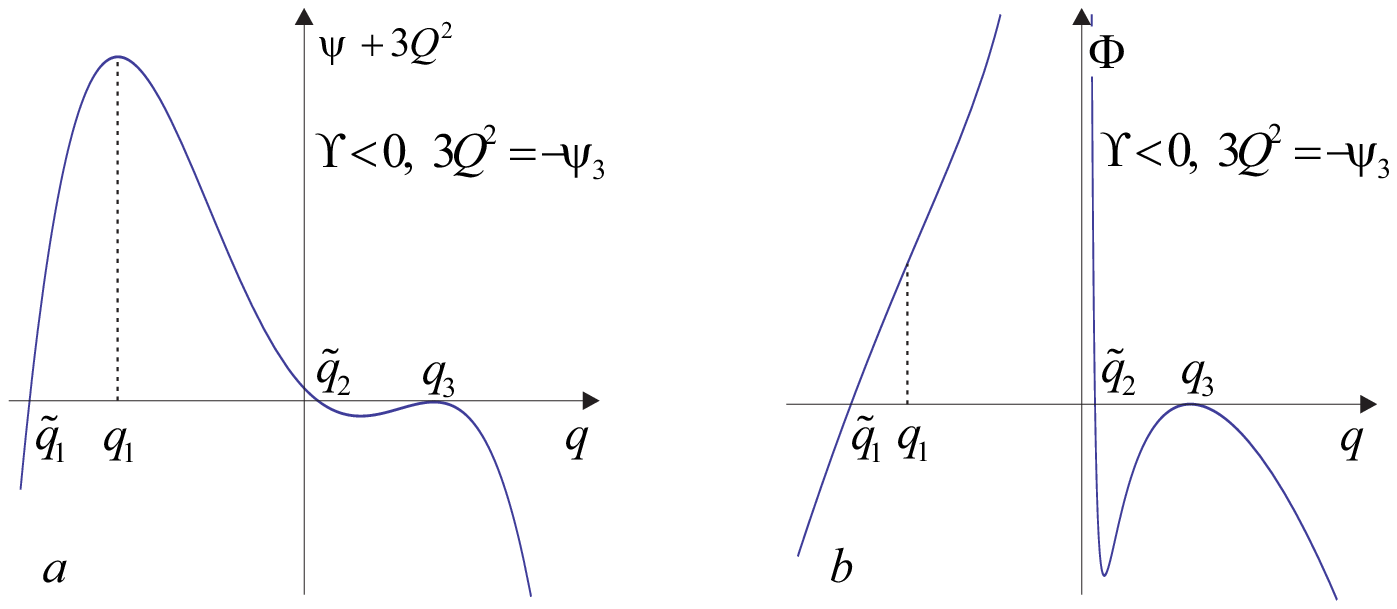}
\hfill {}
\centering\caption{Qualitative behavior of the function $\psi(q)+3Q^2$ and
conformal factor $\Phi(q)$ for $\Lm<0$, $\Upsilon<0$, and $3Q^2=-\psi_3$,
$\psi_2\ne\psi_3<0$.}
\label{f33psiPhi5EM+G}
\end{figure}

Finally, there is one more case for the negative cosmological constant. If
$\psi_2\ne\psi_3<0$, then there arises the possibility:
\begin{equation}                                                  \label{ubvgzr}
  \Lm<0,\qquad\Upsilon<0,\qquad3Q^2=-\psi_3,\qquad\psi_2\ne\psi_3<0.
\end{equation}
In this case, the function $\psi(q)+3Q^2$ and conformal factor are shown in
Fig.~\ref{f33psiPhi5EM+G}
The conformal factor has three zeroes: two simple roots $\tilde q_1$ and
$\tilde q_2$ and one zero $\tilde q_3=q_3$ of the second order. Thus, we have
five global solutions which were already met:
\begin{equation}                                                  \label{ubdnsk}
\begin{aligned}
  (-\infty,\tilde q_1)&:~R8_{(--)}, & \qquad(\tilde q_2,q_3)&:~R10_{(--)},
\\
  (\tilde q_1,0)&:~R4_{(++)}, & (q_3,\infty)&:~R9_{(--)}.
\\
  (0,\tilde q_2)&:~R4_{(++)}, &
\end{aligned}
\end{equation}
\subsubsection{Positive cosmological constant $\Lm>0$}
For positive cosmological constant, the following equality holds
\begin{equation*}
  \Upsilon:=\frac1{8\Lm^3}+\frac{9M^2}{16\Lm^2}>\left(\frac{3M}{4\Lm}\right)^2,
  \qquad\forall M\ge0.
\end{equation*}
Therefore Eq.~(\ref{uvskju}) has only one nonnegative real root
\begin{equation*}
  q_4:=\sqrt[3]{\frac{3M}{4\Lm}+\sqrt\Upsilon}
  +\sqrt[3]{\frac{3M}{4\Lm}-\sqrt\Upsilon}.
\end{equation*}
It corresponds to the minimum of the auxiliary function $\psi_4:=\psi(q_4)<0$.
If
\begin{equation*}
  M=0\qquad\Leftrightarrow\qquad\Upsilon=\frac1{8\Lm^2},
\end{equation*}
then $q_4=0$.

\begin{figure}[hbt]
\hfill\includegraphics[width=.8\textwidth]{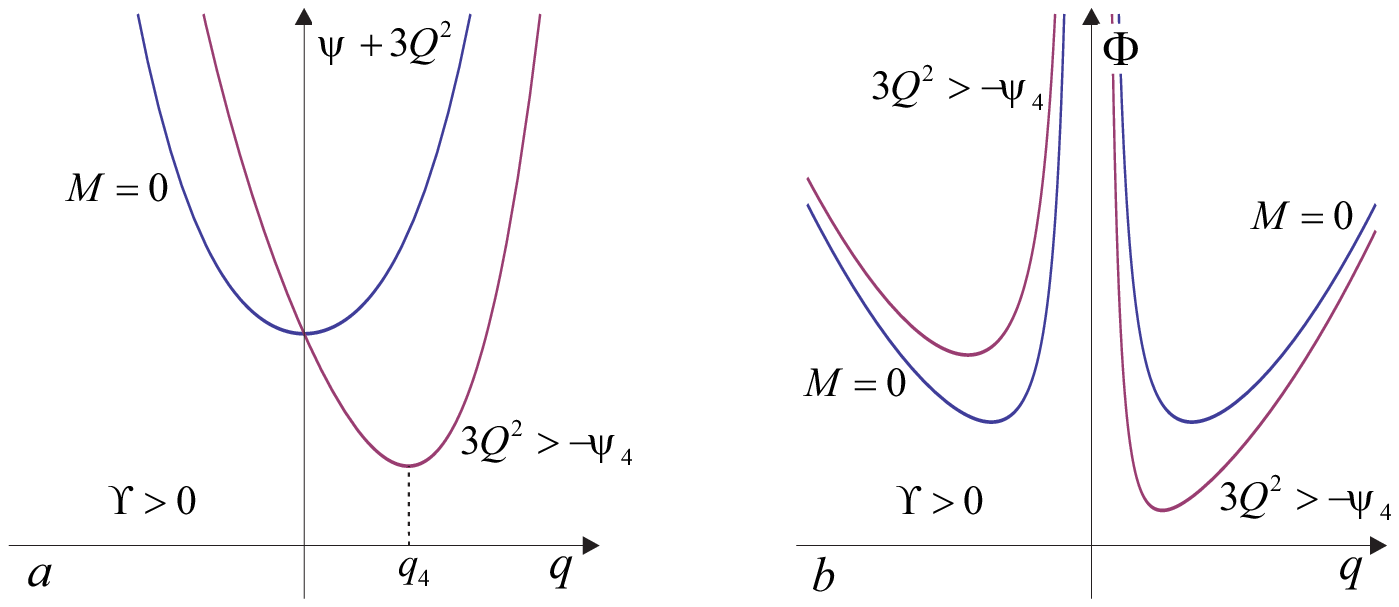}
\hfill {}
\centering\caption{Qualitative behavior of the function $\psi(q)+3Q^2$ and
conformal factor $\Phi(q)$ for $\Lm>0$, $\Upsilon>0$ и $3Q^2>-\psi_4$,
$\psi_2\ne\psi_3<0$.}
\label{f33psiPhi6EM+G}
\end{figure}

The conformal factor does not have zeroes, if the inequalities
\begin{equation}                                                  \label{unbcgt}
\begin{split}
  \Lm>0,\qquad\Upsilon>0,&\qquad M=0,\qquad 3Q^2>0,
\\
  \Lm>0,\qquad\Upsilon>0,&\qquad M>0,\qquad 3Q^2>-\psi_4
\end{split}
\end{equation}
hold. The corresponding functions $\psi(q)+3Q^2$ and conformal factors are shown
in Fig.~\ref{f33psiPhi6EM+G}.
In these cases, there are two global solutions of the same topology related by
the simple reflection $\s\to-\s$:
\begin{equation*}
  (-\infty,0),~(0,\infty):\quad R11_{(++)}.
\end{equation*}
This surface is shown in Fig.~\ref{f31RimSurLmGR+M}. It is geodesically
incomplete at the point $q=0$, where the scalar curvature has singularity. The
length of the directing circle goes to infinity if $q\to\pm\infty$, and the
geodesics are complete there. The corresponding value of $\s$ is finite.

\begin{figure}[hbt]
\hfill\includegraphics[width=.8\textwidth]{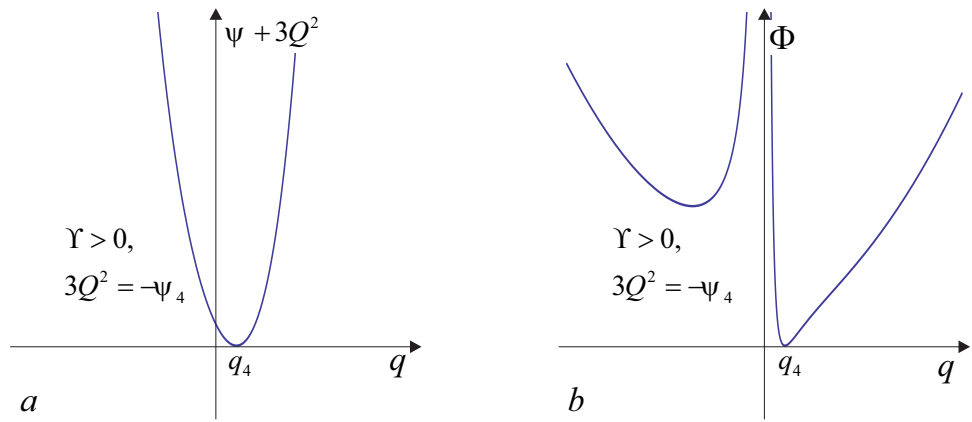}
\hfill {}
\centering\caption{Qualitative behavior of the function $\psi(q)+3Q^2$ and
conformal factor $\Phi(q)$ for $\Lm>0$, $\Upsilon>0$, and $3Q^2=-\psi_4$.}
\label{f33psiPhi7EM+G}
\end{figure}

If
\begin{equation}                                                  \label{unnbcg}
  \Lm>0,\qquad\Upsilon>0,\qquad 3Q^2=-\psi_4,
\end{equation}
then the function $\psi(q)+3Q^2$ has one zero of the second order. This
function and the corresponding conformal factor are shown in
Fig.~\ref{f33psiPhi7EM+G}
There are three already met Riemannian surfaces $\MV$,
\begin{equation}                                                  \label{unbgct}
  (-\infty,0):~R11_{(++)},\qquad(0,q_4):~R2_{(++)},\qquad
  (q_4,\infty):~R9_{(++)},
\end{equation}
if the inequalities (\ref{unnbcg}) hold.

\begin{figure}[hbt]
\hfill\includegraphics[width=.8\textwidth]{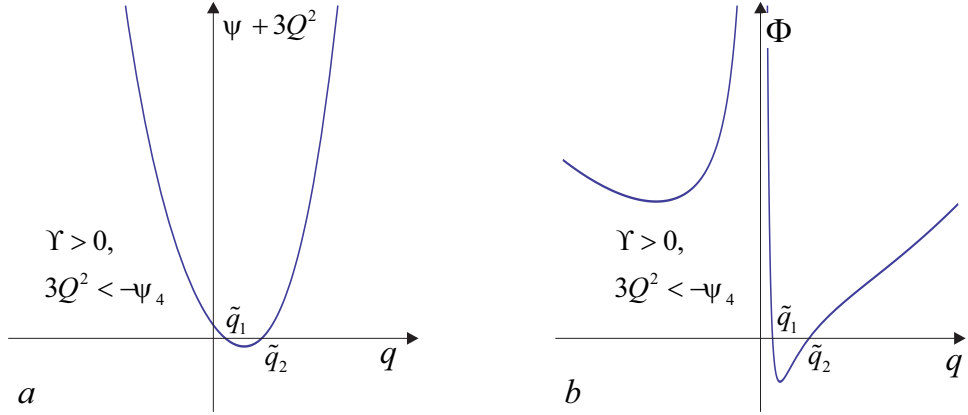}
\hfill {}
\centering\caption{Qualitative behavior of the function and $\psi(q)+3Q^2$ and
conformal factor $\Phi(q)$ for $\Lm>0$, $\Upsilon>0$, and $3Q^2<-\psi_4$.}
\label{f34psiPhi8EM+G}
\end{figure}

The last possibility for positive cosmological constant arises, when
\begin{equation}                                                  \label{unndhy}
  \Lm>0,\qquad\Upsilon>0,\qquad 3Q^2<-\psi_4.
\end{equation}
Graphs of the function $\psi(q)+3Q^2$ and conformal factor are shown in
Fig.~\ref{f34psiPhi8EM+G}.
The conformal factor has two zeroes in this case. Therefore there are four
global solutions:
\begin{equation}                                                  \label{ubvgfr}
\begin{aligned}
  (-\infty,0)&:~R11_{(++)}, &\qquad (\tilde q_1,\tilde q_2)&:
  ~R5_{(--)},~R6_{(--)},
\\
  (0,\tilde q_1)&:~R4_{(++)}, & (\tilde q_2,\infty)&:~R8_{(++)}.
\end{aligned}
\end{equation}
These solutions were already met.
\subsection{Solutions with the Minkowskian plane, $K^{(g)}=0$    \label{speprp}}
The geodesically complete surface $\MU$ for $K^{(g)}=0$ is the Minkowskian
plane $\MR^{1,1}$, or a cylinder, or torus (after compactification). There are
new solutions interesting from the topological point of view. The corresponding
four-dimensional metric in Schwarzschild-like coordinates is
\begin{equation}                                                  \label{elnmps}
  ds^2=q^2(dt^2-dx^2)+\frac{dq^2}{\Phi(q)}+\Phi(q)d\rho^2,
\end{equation}
where
\begin{equation}                                                  \label{uvbcgd}
  \Phi(q)=-\frac{2M}q+\frac{Q^2}{q^2}+\frac{\Lm q^2}3.
\end{equation}
Depending on the values of the constants $\Lm$, $Q$, and $M$ entering the
conformal factor, there are different maximally extended along geodesics
surfaces $\MV$. Let us start with the simplest case.
\subsubsection{Zero cosmological constant $\Lm=0$}
\begin{figure}[hbt]
\hfill\includegraphics[width=.4\textwidth]{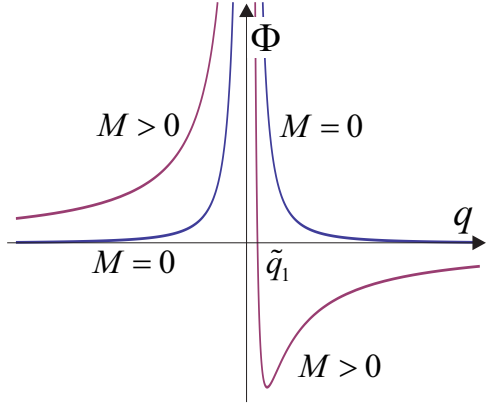}
\hfill {}
\centering\caption{Qualitative behavior of the conformal factor $\Phi(q)$ for
$K^{(g)}=0$, $\Lm=0$, and $M=0$ or $M>0$.}
\label{f34Phi0Lm0EM+G}
\end{figure}

For zero cosmological constant, the conformal factor for the surface $\MV$ is
\begin{equation*}
  \Phi(q)=-\frac{2M}q+\frac{Q^2}{q^2}.
\end{equation*}
Its qualitative behavior for $M=0$ and $M>0$ is shown in
Fig.~\ref{f34Phi0Lm0EM+G}.
If $M=0$, then the conformal factor is the even function, does not have zeroes,
and there are only two identical global solutions, which were already met:
\begin{equation}                                                  \label{eabdfr}
  \Lm=0,\quad M=0:\qquad (-\infty,0),~(0,\infty):\quad R2_{(++)}.
\end{equation}
In contrast to the previous cases, the scalar curvature of the surface $\MV$
tends to zero for $q\to\pm\infty$. If $M>0$, then the conformal factor has one
positive zero $\tilde q_1=Q^2/2M$, and consequently there are three maximally
extended surfaces $\MV$, which were met too:
\begin{equation}                                                  \label{unmhfr}
  (-\infty,0):~R2_{(++)},\qquad(0,\tilde q_1):~R4_{(++)},\qquad
  (\tilde q_1,\infty):~R10_{(--)}.
\end{equation}
In the last case the two-dimensional scalar curvature goes to zero when
$q\to\infty$.
\subsubsection{Negative cosmological constant $\Lm<0$}
\begin{figure}[hbt]
\hfill\includegraphics[width=.8\textwidth]{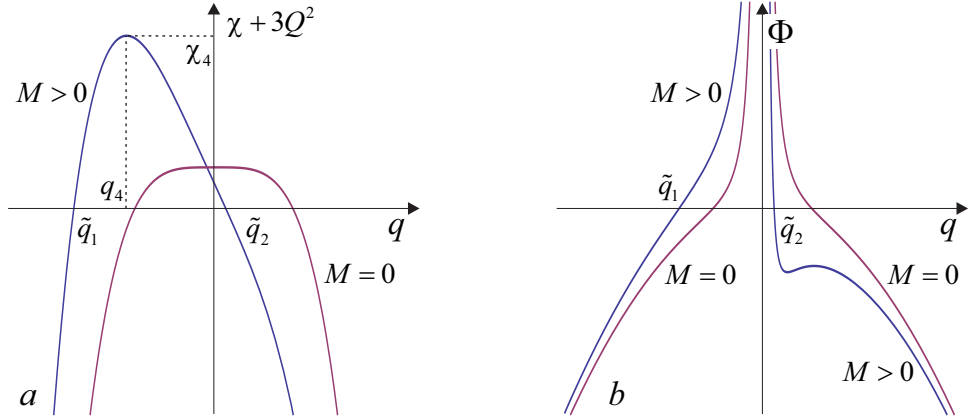}
\hfill {}
\centering\caption{Qualitative behavior of the auxiliary function $\chi(q)+3Q^2$
and conformal factor $\Phi(q)$ for $K^{(g)}=0$, $\Lm<0$ for $M=0$ and $M>0$.}
\label{f34chiPhi1abEM+G}
\end{figure}

To analyze the conformal factor (\ref{uvbcgd}) for nonzero cosmological
constant we write the conformal factor as
\begin{equation*}
  \Phi(q)=-\frac{2M}q+\frac{Q^2}{q^2}+\frac{\Lm q^2}3=\frac{\chi(q)+3Q^2}{3q^2},
\end{equation*}
where the auxiliary function is introduced
\begin{equation}                                                  \label{ujsdgg}
  \chi(q):=\Lm q^4-6Mq=q(\Lm q^3-6M).
\end{equation}
The constant (\ref{ebgtdj}) is positive for solutions of the cubic equation
$\chi'=0$, defining locations of extrema of the auxiliary function,
\begin{equation*}
  \Upsilon=\frac{9M^2}{16\Lm^2}>0.
\end{equation*}
Consequently, the auxiliary function $\chi(q)$ has one extremum at the point
\begin{equation*}
  q_4=\sqrt[3]{\frac{3M}{2\Lm}}
\end{equation*}
with the corresponding value
\begin{equation}                                                  \label{uvxbdr}
  \chi_4:=\chi(q_4)=-\frac{9M}2\sqrt[3]{\frac{3M}{2\Lm}}.
\end{equation}
Since the value $\chi_4>0$ is positive for $\Lm<0$, and the graph of the
auxiliary function is shifted upwards on $3Q^2$, the conformal factor has
only two simple zeroes. Qualitative behavior of the function $\chi(q)+3Q^2$ and
conformal factor are shown in Fig.~\ref{f34chiPhi1abEM+G}.
From topological standpoint, maximally extended surfaces $\MV$ for $M=0$ and
$M>0$ coinside. There are only two different surfaces:
\begin{equation}                                                  \label{unbdfg}
  (-\infty,\tilde q_1),~(\tilde q_2,\infty):~R8_{(--)},\qquad
  (\tilde q_1,0),~(0,\tilde q_2):~R4_{(++)},
\end{equation}
which were already met.
\subsubsection{Positive cosmological constant $\Lm>0$}
If cosmological constant is positive and $M=0$, then the auxiliary function
$\chi(q)$ has one zero $q=0$. In this case, the conformal factor is an even
function, does not have zeroes for $Q>0$, and there are two identical global
solutions:
\begin{equation}                                                  \label{esgwht}
  (-\infty,0),~(0,\infty):\qquad R2_{(++)}.
\end{equation}
The difference from the previous cases is that the scalar curvature $R\to0$ for
$q\to\pm\infty$.

\begin{figure}[hbt]
\hfill\includegraphics[width=.8\textwidth]{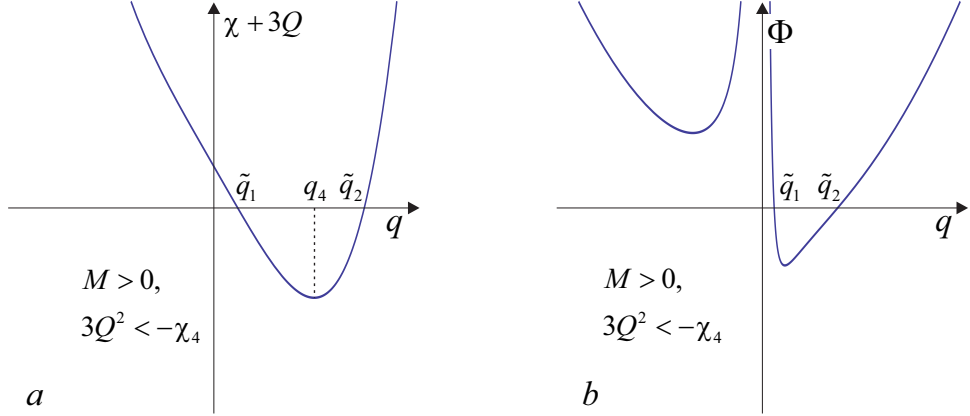}
\hfill {}
\centering\caption{Qualitative behavior of the function $\chi(q)+3Q^2$ and
conformal factor $\Phi(q)$ for $K^{(g)}=0$, $\Lm>0$ when $M>0$ and
$3Q^2<-\chi_4$.}
\label{f34chiPhi2abEM+G}
\end{figure}

For positive $M>0$, the auxiliary function $\chi(q)$ has one minimum for
$q_4>0$. In addition, $\chi_4<0$. If $3Q^2<-\chi_4$, then the conformal factor
has two positive simple zeroes. Qualitative behavior of the function
$\chi(q)+3Q^2$ and conformal factor $\Phi(q)$ for this case are shown in
Fig.~\ref{f34chiPhi2abEM+G}.
The conformal factor has two positive simple zeroes at points $\tilde q_{1,2}$,
and consequently there are four global solutions:
\begin{equation}                                                  \label{eqwtre}
\begin{aligned}
  (-\infty,0):&~R11_{(++)}, &  (\tilde q_1,\tilde q_2):&~R5_{(--)},~R6_{(--)},
\\
  (0,\tilde q_1):&~R4_{(++)},\qquad    & (\tilde q_2,\infty):&~R8_{(++)}.
\end{aligned}
\end{equation}
Topologically equivalent solutions were already met.

\begin{figure}[hbt]
\hfill\includegraphics[width=.8\textwidth]{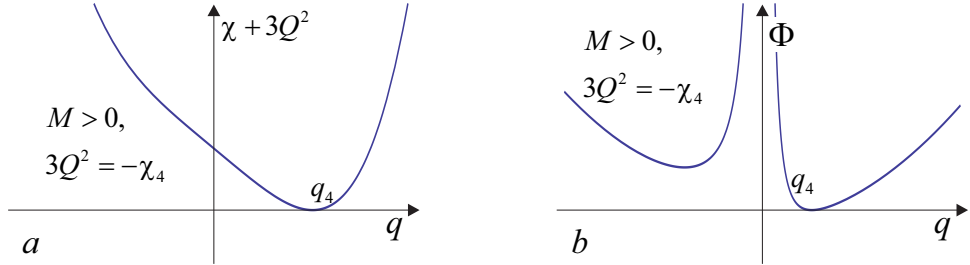}
\hfill {}
\centering\caption{Qualitative behavior of the function $\chi(q)+3Q^2$ and
conformal factor $\Phi(q)$ when $K^{(g)}=0$ and $\Lm>0$ for $M>0$ and
$3Q^2=-\chi_4$.}
\label{f34chiPhi3abEM+G}
\end{figure}

If $3Q^2=-\chi_4$, then the conformal factor has one second order zero at
point $q_4$. The corresponding function $\chi(q)+3Q^2$ and conformal factor
$\Phi(q)$ are shown in Fig.~\ref{f34chiPhi3abEM+G}.
In this case, the conformal factor has one positive root of the second order.
Therefore there are three maximally extended surfaces $\MV$:
\begin{equation}                                                  \label{efnhtu}
  (-\infty,0):~R11_{(++)},\qquad (0,q_4):~R2_{(++)},\qquad
  (q_4,\infty):~R9_{(++)}.
\end{equation}
We note the behavior of the two-dimensional scalar curvature: $R\to0$ when
$q\to\pm\infty$ and $R\to\const$ when $q\to q_4$.

If the inequality $3Q^2>-\chi_4$ holds, then the conformal factor does not have
zeroes, and there are only two topologically equivalent global solutions:
\begin{equation}                                                  \label{eagsth}
  (-\infty,0),~(0,\infty):\qquad R11_{(++)}.
\end{equation}

Thus, we have constructed all global solutions of the form $\MR^{1,1}\times\MV$
with metric (\ref{elnmps}). Topologically all surfaces $\MV$ were already met in
the case of Lorentz invariant solutions of the form $\ML^2\times\MV$, but now
the first factor is different. Therefore there are 8 new topologically
inequivalent solutions which are listed in Eqs.~(\ref{eabdfr}), (\ref{unmhfr}),
(\ref{unbdfg}), (\ref{esgwht}), (\ref{eqwtre}), (\ref{efnhtu}), and
(\ref{eagsth}).

The unexpected property of the maximally extended surfaces $\MV$ (at least for
the authors) is the following. Suppose that the conformal factor $\Phi(q)$ is
defined on the whole real line $q\in\MR$, where it has finite number of zeroes
and singularities. Then every maximally extended surface $\MV$ corresponds to
one of the intervals between neighboring zeroes and singularities. Inversely,
there is one maximally extended surface for each interval. As a result, there
are several different global solutions of the form $\MR^{1,1}\times\MV_\Si$,
where index $\Si$ enumerates intervals, corresponding to one solution of the
field equations. The metric on these solutions may have different
signatures, depending on the sign of the conformal factor. So, for fixed
coupling constants in the action, Einstein's equations admit both signature
metrics, and they are different.
\section{Conclusion}
We have found and classified all global solutions of Einstein's equations
with a cosmological constant and electromagnetic field which have the form
of the warped product of two surfaces, $\MM=\MU\times\MV$. The equations of
motions imply that at least one of the surfaces must be of constant curvature.
There are only three cases (\ref{ecasek}). The cases {\sf A} and {\sf B} were
considered in \cite{AfaKat19}. In the present paper we classified solutions in
the case {\sf C}, when the Lorentzian surface $\MU$ is of constant curvature.

In the case {\sf A}, when both surfaces are of constant curvature, there are
6 Killing vector fields. In each of the cases {\sf B} and {\sf C}, there are 4
Killing vector fields: the constant curvature surface has three Killing vectors
and the other surface has one (the analog of Birkhoff's theorem). All solutions
in the case {\sf C} are invariant under the Lorentz group $\MS\MO(1,2)$
(Lorentz invariant solutions) or the Poincar\'e group $\MI\MO(1,1)$ (planar
solutions) acting on surface $\MU$. This phenomenon is called the spontaneous
symmetry emergence.

Solutions were explicitly constructed for all symmetry groups and all values of
cosmological constant $\Lm$, charge $Q$, and constant of integration $M$, which
has the meaning of mass for the Schwarzschild solution. We have found 19 global
solutions. The question whether there are topologically equivalent among them is
open for future research.
We see that the requirement of maximal
extension of solutions along geodesics defines almost uniquely the global
structure of space-time. It is important that we do not use any boundary
conditions on the solutions.

Let us mention the important property. Assume that the sign of the
electromagnetic part of the action is fixed by the requirement of positive
definiteness of the canonical Hamiltonian for physical degrees of freedom for
metric signature $(+---)$. Einstein's equations are such that for some values of
the cosmological constant $\Lm$, charge $Q$, and integration constant $M$ there
are global solutions with metrics of both signatures $(+---)$ and $(-+++)$. That
is for fixed signs in the action there are topologically disconnected solutions
with and without ghosts. It seems, that solutions with ghosts must be discarded
as unphysical. It is important that this cannot be achieved by choosing the
signs in the action.

Knowledge of global structure of the space-time allows one to give physical
interpretation of solutions. We showed that solutions with the electromagnetic
field describe black holes, naked singularities, cosmic strings, wormholes,
domain walls of curvature singularities and many others. In the present paper,
we only briefly discussed their physical properties.


\end{document}